\documentclass[a4paper]{article}
\usepackage[margin=20mm]{geometry}
\usepackage{amsmath}
\usepackage{amsfonts}
\usepackage{amssymb}
\usepackage{graphicx}
\usepackage{setspace}
\usepackage{lineno}
\usepackage{verbatim}
\usepackage{comment}
\usepackage{soul}
\immediate\write18{texcount -tex -sum  \jobname.tex > \jobname.wordcount.tex}

\usepackage{natbib}
\bibpunct{(}{)}{;}{a}{}{,} 

\usepackage{txfonts}
\usepackage{tabularx}
\usepackage{hyperref}
\providecommand{\keywords}[1]
{
  \small	
  \textbf{\textit{Keywords---}} #1
}

   \title{The five largest satellites of Uranus: astrometric observations spread
   over 29 years at the Pico dos Dias Observatory
   }

   \author{
Camargo, J.I.B.$^{1,2}$,
Veiga, C.H.$^{1}$,
Vieira-Martins, R.$^{1,2}$,
Fienga, A.$^{3,4}$,
Assafin, M.$^{5,2}$\\

\small $^{1}$Observatório Nacional/MCTI, R. General José Cristino 77, CEP 20921-400 Rio de Janeiro, RJ, Brazil\\
\small $^{2}$Laboratório Interinstitucional de e-Astronomia - LIneA, Rua Gal. José Cristino 77, Rio de Janeiro, RJ 20921-400, Brazil\\
\small $^{3}$Géoazur-CNRS 7329, Observatoire de la Côte d’Azur, Valbonne, France\\
\small $^{4}$IMCCE-CNRS 8028, Observatoire de Paris, Paris, France\\
\small $^{5}$Observatório do Valongo/UFRJ, Ladeira do Pedro Antônio 43, CEP 20080-090 Rio de Janeiro, RJ, Brazil
             }

   \date{Received:; Accepted:}

\begin{document} 

  \maketitle

%
  \begin{abstract}
   We present the astrometry of the five largest satellites of Uranus from 
observations spread over almost three decades with photographic plates and 
CCDs (mainly), taken at the Pico dos Dias Observatory - Brazil. All positions 
presented here are obtained from the reanalysis of measurements and images 
used in previous publications. Reference stars are those from the Gaia Early 
Data Release 3 (Gaia EDR3) allowing, in addition to a higher accuracy, a larger
number of positions of the largest satellites as compared to our previous works. 
From 1982 to 1987, positions were obtained from photographic plates. From 1989 
to 2011, CCDs were used. On average, we obtained 
$\Delta\alpha{\rm cos}\delta=-11~(\pm52)$ milli-arcseconds and 
$\Delta\delta=-14~(\pm43)$ milli-arcseconds for the differences in the sense 
observation minus ephemerides (DE435$+$ura111). Comparisons with different 
ephemerides (DE440, INPOP21a, INPOP19a and NOE-7-2013-MAIN) and results from 
stellar occultations indicate a possible offset in the (Solar System) 
barycentric position of the Uranian system barycenter. Overall, our results 
are useful to improve dynamical models of the Uranian largest satellites as 
well as the orbit of Uranus.

   \end{abstract}


\section{Introduction}

Seventh planet from the Sun, the ice giant Uranus was discovered by accident by Sir William Herschel (1738-1822) in 1781 \citep{2009gpss.book.....I}. Few years later, in 1787, Herschel himself discovered Uranus' two largest moons: Titania and Oberon. Ariel and Umbriel were discovered later, in 1851, by William Lassell. Miranda, completing the five largest satellites of Uranus, was discovered almost a century later, in 1948, by Gerard Kuiper \citep{2016mss..book.....H}. Table~\ref{tab:satcharacteristics} presents relevant information about the main satellites and a review of the dynamics of the Uranian system, as well as its observations, can be seen in \citet{2013MNRAS.436.3668E} and \citet{2014AJ....148...76J}.

Astrometry of these satellites are relevant mainly to improve their dynamical models \citep[see, for instance, ][]{1987A&A...188..212L, 2016CeMDA.126..145L} as well as the orbit of Uranus itself \citep[][- INPOP10e and INPOP21a respectively]{2013arXiv1301.1510F, INPOP21A}.
It should be clarified here, in addition, the context of the works by  \citet{2013MNRAS.436.3668E} and \citet{1987A&A...188..212L}. The first presents an example of ephemeris built on the basis of observations spread 
over a time interval of 220 years and reaching dates close to those of the discovery of the satellites. The latter was based on a general analytical theory \citep[GUST - General Uranus Satellite Theory,][]{1986A&A...166..349L}, fitted to Earth-based observations acquired from 1911 to 1986 and to optical navigation and radio data from Voyager.

Positions of these satellites may also be improved from mutual events \citep[e.g.][]{2008MNRAS.384L..38H,2009AJ....137.4046A,2013A&A...557A...4A} and stellar occultations \citep[e.g.][]{2009Icar..199..458W}. In addition, with this latter technique, sizes/shapes can be determined and the presence of rings, atmospheres and even topographic features can be detected and studied \citep[see, for instance, ][for some examples involving different objects]{2011Natur.478..493S,2012Natur.491..566O,2013ApJ...773...26B,2014Natur.508...72B,2015MNRAS.451.2295G,2015ApJ...811...53D,2016ApJ...819L..38S,2017AJ....154...22D,2017Natur.550..219O}. Whatever the case, accurate astrometry of Solar System objects is needed. 

This paper presents an homogeneous analysis to translate plate/CCD coordinates into celestial equatorial coordinates from an observational data set explored by \citet{2003AJ....125.2714V} and \citet{2015A&A...582A...8C} (C15 hereafter). This data set (plate/CCD measurements and images, see details later in the text) represents the observational history of our team from 1982 to 2011 associated to the astrometry of the main satellites of Uranus (exception made to the observations of mutual phenomena). The main improvements over our previous publications are: a larger number of positions of the main Uranian satellites; use of Gaia EDR3 as reference for astrometry (therefore, a larger number of reference stars and virtually no systematics due to the reference catalogue); same procedures to translate plate/CCD coordinates into celestial equatorial coordinates in the ICRF~\citep[International Celestial Reference Frame,][]{1998AJ....116..516M}. In brief, more than two decades of observations of the Uranus' main satellites treated in a homogeneous way, providing accurate astrometry along with a better and more complete use of our observational data set thanks to Gaia EDR3. The coronographed images used in C15 were completely remeasured.

In Sect. 2, we briefly present the data and the instruments from which they were obtained. Methods and a brief discussion on lower accuracy limits to the determination of centroids are presented in Sect. 3 and the data analysis, in Sect. 4. Conclusions and comments are then provided in Sect. 5.

\section{Observations and data}

All observations used in this paper were made at the Pico dos Dias Observatory (IAU code: 874, run by the Laboratório Nacional de Astrofísica/MCTI\footnote{\url{https://www.gov.br/mcti/pt-br/rede-mcti/lna}}), using photographic plates (MAY/1982 to JUL/1987) and CCDs (SEP/1989 to OCT/2011) and involving telescopes of apertures 1.6m (Perkin-Elmer - photographic plates, CCDs) and 0.6m (Boller\&Chivens or Zeiss - CCDs only). A brief description of the photographic plates (Table~\ref{tab:plates}) as well as the distribution of observations per night (Fig.~\ref{fig:histobs}) are given below. Further details about the instruments can be found in \citet{1987A&AS...70..325V}, \citet{2003AJ....125.2714V} and C15.

\begin{table}
\small
\centering
\caption[Characteristics of the satellites.]{Columns: satellite ID; orbital inclination; semi-major axis; satellite size; sidereal orbit period; calculated magnitude. {\it a}, {\it i} and {\it T} were obtained from the JPL HORIZONS Web-Interface, using as reference epoch J2000 TDB, the Uranus equator and node of date as reference plane,
and Uranus as the central body. Magnitudes were also obtained from the JPL HORIZONS Web-Interface. Mean radii were obtained from the JPL's Planetary Satellite Physical Parameters. The reference mentioned therein for them is \citet{2018CeMDA.130...22A}. Ephemeris source: ura111.}
\begin{tabular}{cccccc}
\hline
Satellite & \it{i} & \it{a}      & Mean radius   & \it{T} & Mag. \\
          & (deg.) & (km)        & (km)          & (days) & (V)  \\
\hline
Miranda   & 175.6 & 129\,871.8 &      235.8      &  1.414 & 16.7 \\
Ariel     & 180.0 & 190\,941.3 &      578.9      &  2.521 & 14.5 \\
Umbriel   & 180.0 & 266\,012.3 &      584.7      &  4.145 & 15.2 \\
Titania   & 179.9 & 436\,294.5 &      788.9      &  8.706 & 14.1 \\
Oberon    & 179.8 & 583\,551.9 &      761.4      & 13.468 & 14.3 \\
\hline
\end{tabular}
\label{tab:satcharacteristics}
\end{table}

\begin{table}
\centering
\caption[Photographic plates.]{Characteristics of the photographic plates
as provided by \citet{1987A&AS...70..325V}.}
\begin{tabular}{ccc}
\hline
Manufacturer & Emulsion & Size \\
 & & (cm) \\
\hline
 Kodak & IIIaJ  & 12 $\times$10  \\
 Kodak & IIaO   & 12 $\times$10  \\
 Kodak & IIaD   & 12 $\times$10  \\
 Kodak & 103aO  & 12 $\times$10  \\
\hline
\end{tabular}
\label{tab:plates}
\end{table}

\begin{figure}
\caption{Number of observations per night. This figure takes into consideration all runs, irrespective of the quality of the data and its presence in the final catalogue.}
\centerline{\includegraphics[scale=0.6]{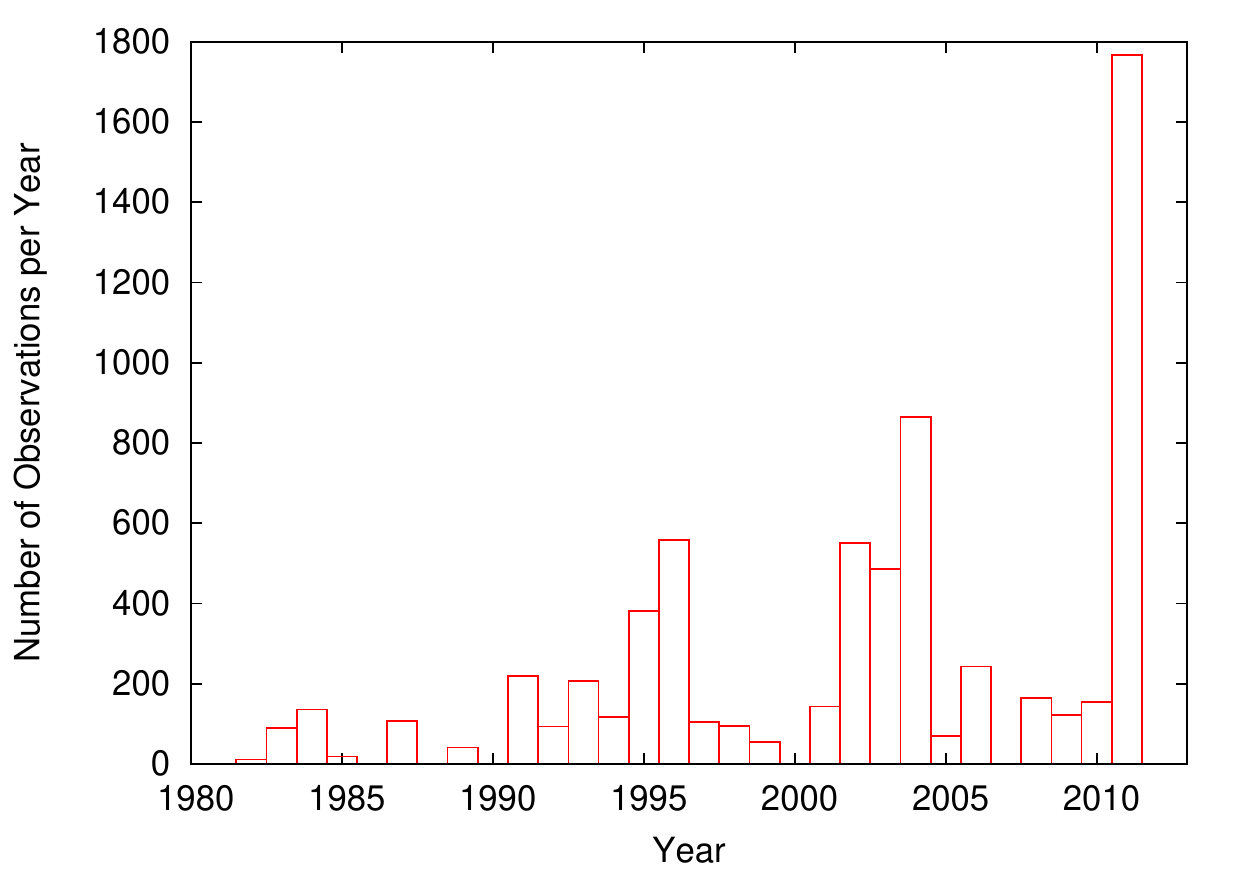}}
\label{fig:histobs}
\end{figure}

\section{Methods and precision limits}

As previously indicated, this paper brings no observations other than those belonging to the observational history of \citet{2003AJ....125.2714V} and C15. Data from the first paper were made available to this work from tables with (x,y) plate/CCD coordinates, mid-time of the observation, astrometric ICRF equatorial coordinates associated to the (x,y) coordinates as well as an identification of the type of object in each of them (field star, Ariel, Miranda, Oberon, Titania, Umbriel and Uranus). Those equatorial coordinates are a first step to the determination of plate or CCD constants. Data from the second paper were obtained directly from images that were submitted to a process of digital coronagraphy \citep[see][C15]{2008P&SS...56.1882A}. Those coronagraphed images were completely re-reduced with the PRAIA astrometric package \citep{2011gfun.conf...85A}.

In all cases, positions from the Gaia EDR3 \citep[Gaia Early Data Release 3,][]{2021A&A...649A...2L} were used as reference for astrometry. The relationship between CCD and gnomonic coordinates were obtained through a first degree polynomial and no Gaia EDR3 star fainter than $G = 16.5$ was used. With this, we avoided using low SNR stars as astrometric references.

CCD observations from \citet{2003AJ....125.2714V} and C15 have an overlap. Whenever it happened, only the results obtained from the image re-reduction were kept. Table~\ref{tab:timeobsinterval} shows the time intervals in which each set of data was taken.

Subroutines from the SPICE toolkit \citep{1996P&SS...44...65A, 2018P&SS..150....9A}, Standards of Fundamental Astronomy\footnote{\url{http://www.iausofa.org}} (SOFA) and Naval Observatory Vector Astrometry Software\footnote{\url{http://aa.usno.navy.mil/software/novas/novas_info. php}} (NOVAS) were used to determine topocentric positions of the satellites and/or to determine geocentric positions, proper motion corrected, reference (Gaia EDR3) stars.

\subsection{Position filtering}

This paper adopts a simpler filtering method as compared to the procedure presented by C15: an iterative $3\sigma$ filter was applied to each satellite in each night it was observed; then, a second iterative $3\sigma$ filter was applied to all observations of each satellite. This was done separately to plate and CCD observations, given the larger standard deviation in the measurements of the former. One of the consequences from this simpler filtering method, thanks to the use of Gaia EDR3 as astrometric reference, was a more efficient exploration of our observational data set and the determination of a larger and more accurate number of CCD positions as compared to \citet{2003AJ....125.2714V} and C15.

One relevant point is the angular distance, projected on to plane of the sky, between Uranus and each satellite. We performed no selection based on it and a discussion is presented later in the text. Our catalogue with positions of the satellites also contains those angular distances along the RA and DEC axes. However, we eliminated by hand from our catalogues 3 positions of Miranda and 1 position of Ariel that were obtained from spurious detections.

\subsection{Position precision}

One interesting verification here is an estimation of the best positional precision we can obtain from the observational data. For this, we will accept the existence of a function ($\Phi$), in the sense of equations (8, 17, 21) in \citet{1983PASP...95..163K}, as a good representation of the flux distribution of the object over the CCD. This is reasonable since the Point Spread Function (PSF) is mostly dominated by the atmosphere and pixel scale is a fraction (10\%-25\%) of the seeing. 

Now, two extreme cases are considered: (i) precision derived from signal dominated signal-to-noise ratios ($\sigma_{\rm bright}$) and (ii) precision derived from sky dominated signal-to-noise ratios. ($\sigma_{\rm faint}$).

Expressions for these two cases, when a circular Gaussian is chosen to describe the PSF (model used in this work), can be given by
equations Eqs.~\ref{eq:sbright} and~\ref{eq:sfaint} (\citet{2005MNRAS.361..861M}, see also \citet{1983PASP...95..163K})
\begin{equation}
  \sigma_{\rm bright}=\frac{\sigma_{\it PSF}}{\sqrt{c}}
  \label{eq:sbright}
\end{equation}
and
\begin{equation}
  \sigma_{\rm faint}=\frac{2\sqrt{2\pi\cdot B}\cdot\sigma^{2}_{\it PSF}}{c}
  \label{eq:sfaint}
\end{equation}, 
where $\sigma_{\it PSF}$ is the standard deviation of the gaussian distribution, {\it c} is the number of electrons due to the source and {\it B} is the number of electrons due to the sky. It should be emphasized that Eqs.~\ref{eq:sbright} and~\ref{eq:sfaint} provide, therefore, the best precision for the object's centroid along the CCD's {\it x}-axis (and, by symmetry, along the {\it y}-axis) in the context of cases (i) and (ii), respectively.

We adopt, from \citet{2005MNRAS.361..861M}, the following expression to determine the lower limit of the centroid uncertainty from a PSF fitting:
\begin{equation}
    \sigma_{x,y}=\sqrt{\sigma^{2}_{\rm bright}+\sigma^{2}_{\rm faint}}
    \label{eq:sfinal}
\end{equation}

\begin{figure}
\caption{The main satellites of Uranus in an image that went through a digital
coronography procedure \citep[see][]{2008P&SS...56.1882A}. North is up, east is left. Image obtained on the night 18-19/JUL/1992 at the Pico dos Dias Observatory with its 1.6m telescope.}
\centerline{\includegraphics[scale=0.7]{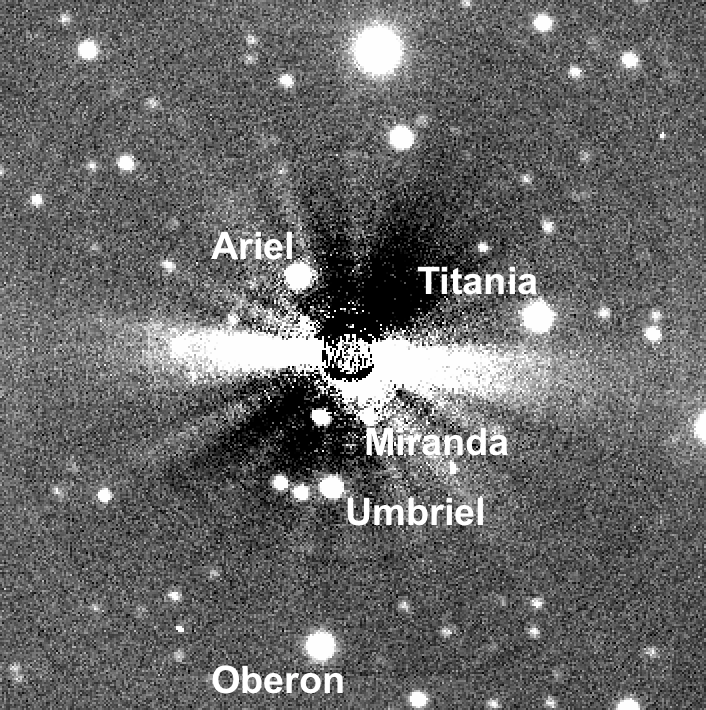}}
\label{fig:urasyst}
\end{figure}

We used images of Ariel and Oberon taken with the 1.6m telescope on 19/JUL/1992 and determined lower limits for their precisions, from Eqs.~\ref{eq:sbright} to \ref{eq:sfinal}, of 5 mas (most frequent value) for Oberon and of 9 mas (most frequent value) for Ariel (see Fig.~\ref{fig:urasyst}). From the differences with the ephemerides, we have ($\sigma_{\alpha}{\rm cos}\delta$, $\sigma_{\delta}$)$=$(13 mas,10 mas) for Oberon and ($\sigma_{\alpha}{\rm cos}\delta$, $\sigma_{\delta}$)$=$(16 mas,11 mas) for Ariel. The determination of {\it B} in Eq.~\ref{eq:sfaint} was obtained from the standard deviation of the background ($\sigma_{\rm bg}$). More specifically, to estimate the background noise overall contribution, we adopted the simple relationship given by
\begin{equation}
    B=n\sigma_{\rm bg}^2
    \label{eq:background}
\end{equation}, 
where {\it n} is the number of background pixels over which the object is sampled.
We opted for this way to estimate the background because the digital coronography procedure subtracts it in the process of attenuation of the scattered light from Uranus. Equation~\ref{eq:background} takes into consideration noise sources other than those from the sky only (the scattered light from Uranus is a relevant such other source). For considerations on the analysis of sky-subtracted CCD data, see \citet{1991PASP..103..122N}.

These lower limits for the precision in the centroid of the Ariel and Oberon are not much smaller than those obtained for their respective right ascension and declination. It is also interesting to mention that, to the images considered in these calculations, the standard deviations from the observed minus calculated values for the reference stars are 22 mas in RA and 14 mas in DEC.

There is an obvious loss of precision when we go from CCD to celestial equatorial coordinates. Possible origins are remaining distortion patterns of the CCD and a PSF that may differ from a circular one. Other issues, mentioned in C15 and that justifies larger standard deviations in RA as compared to those in DEC, are known mechanical problems in the telescope tracking system and incorrect timing inserted in the image header (known to have happened before 2000). Also, as mentioned in that paper, mechanical and timing issues may slightly increase or decrease our right ascensions so that no general systematic effects on the positions of the satellites are expected from them.

\begin{figure*}
\caption{Differences in right ascension (upper panels) and declination (bottom panels) in the sense observations minus ephemerides (DE435+ura111) for Miranda, as a function of time
(left panels), true anomaly (middle panels) and the angle Sun-Observer-Target (right panels). The "*" means multiplication by the cosine of the declination.}
{\includegraphics[scale=0.5]{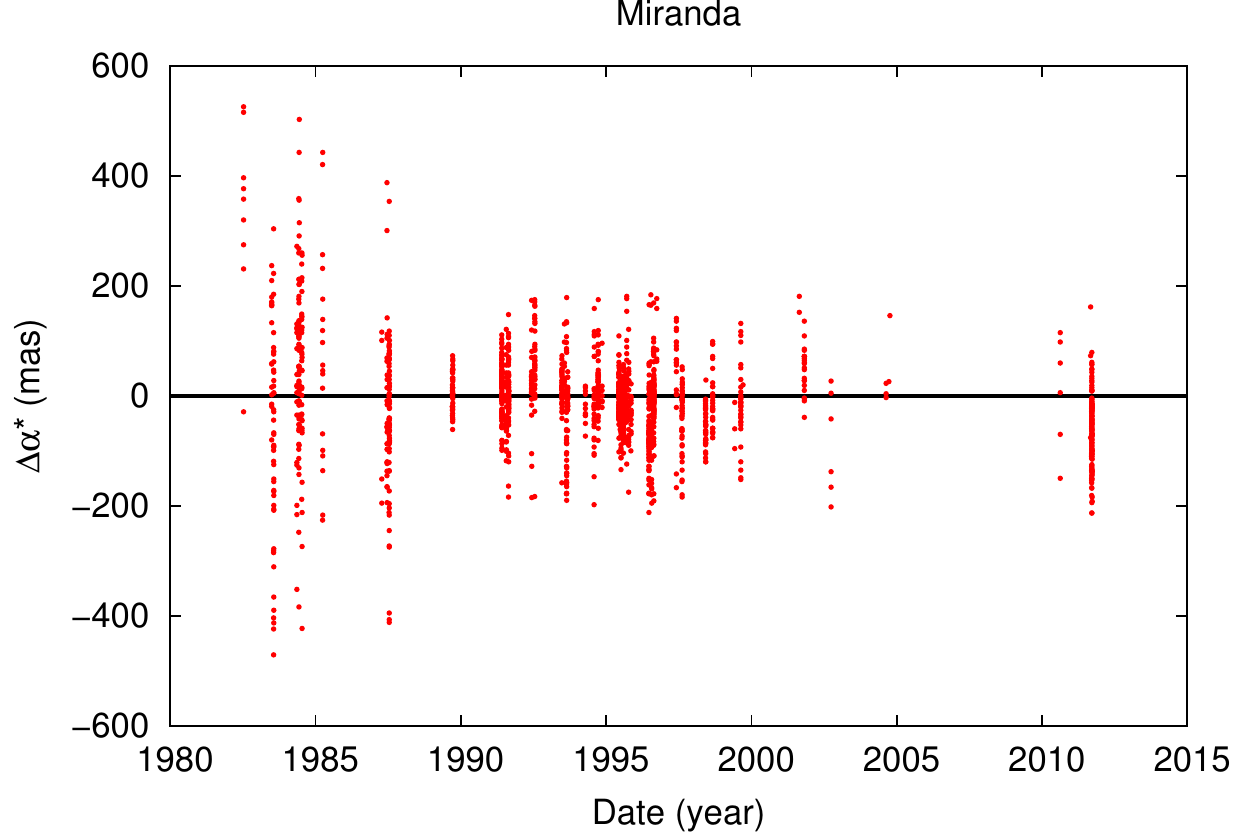}\includegraphics[scale=0.5]{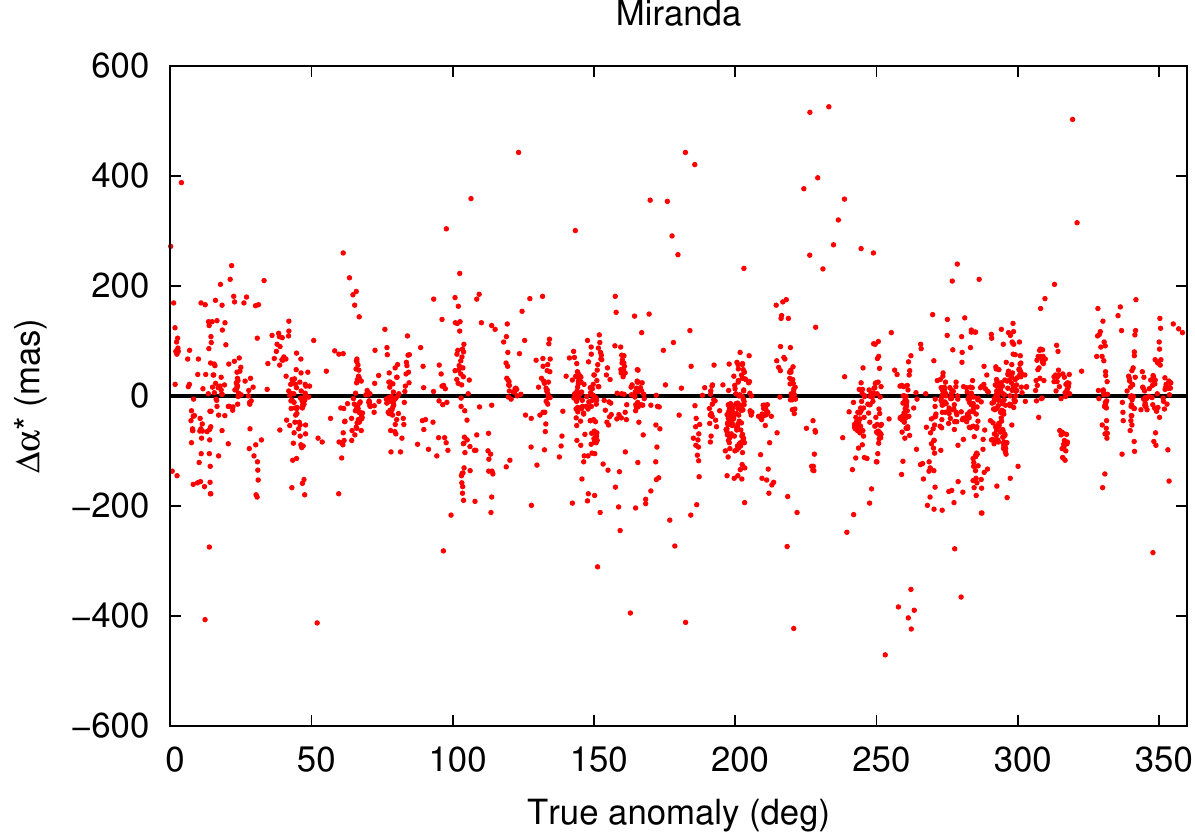}\includegraphics[scale=0.5]{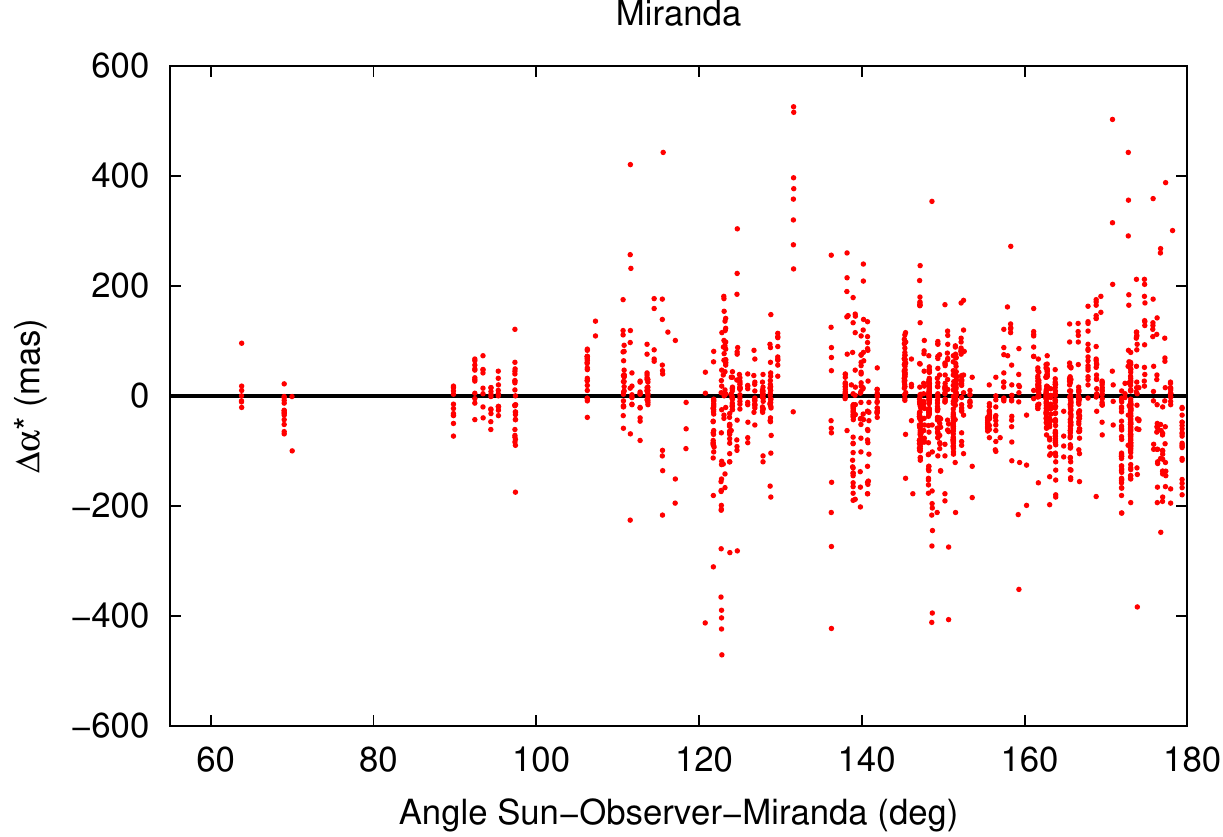}}

{\includegraphics[scale=0.5]{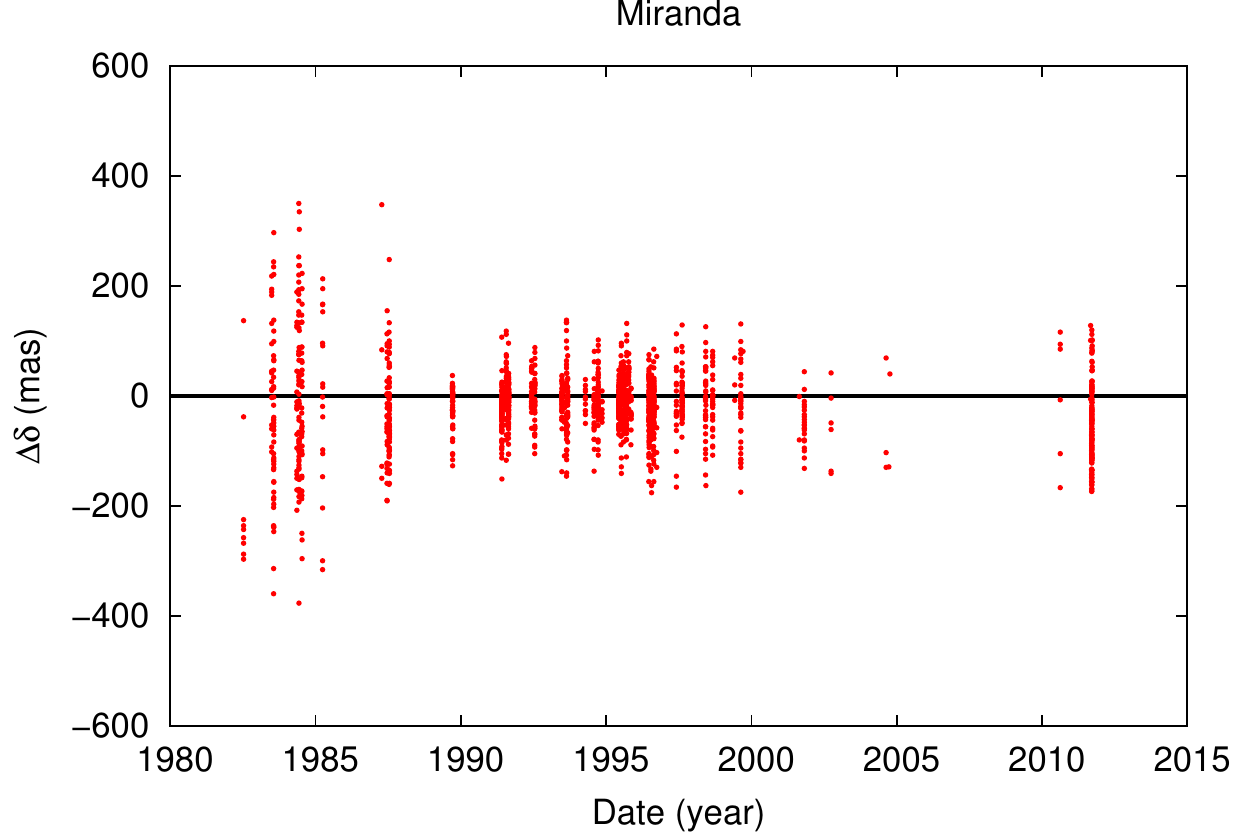}\includegraphics[scale=0.5]{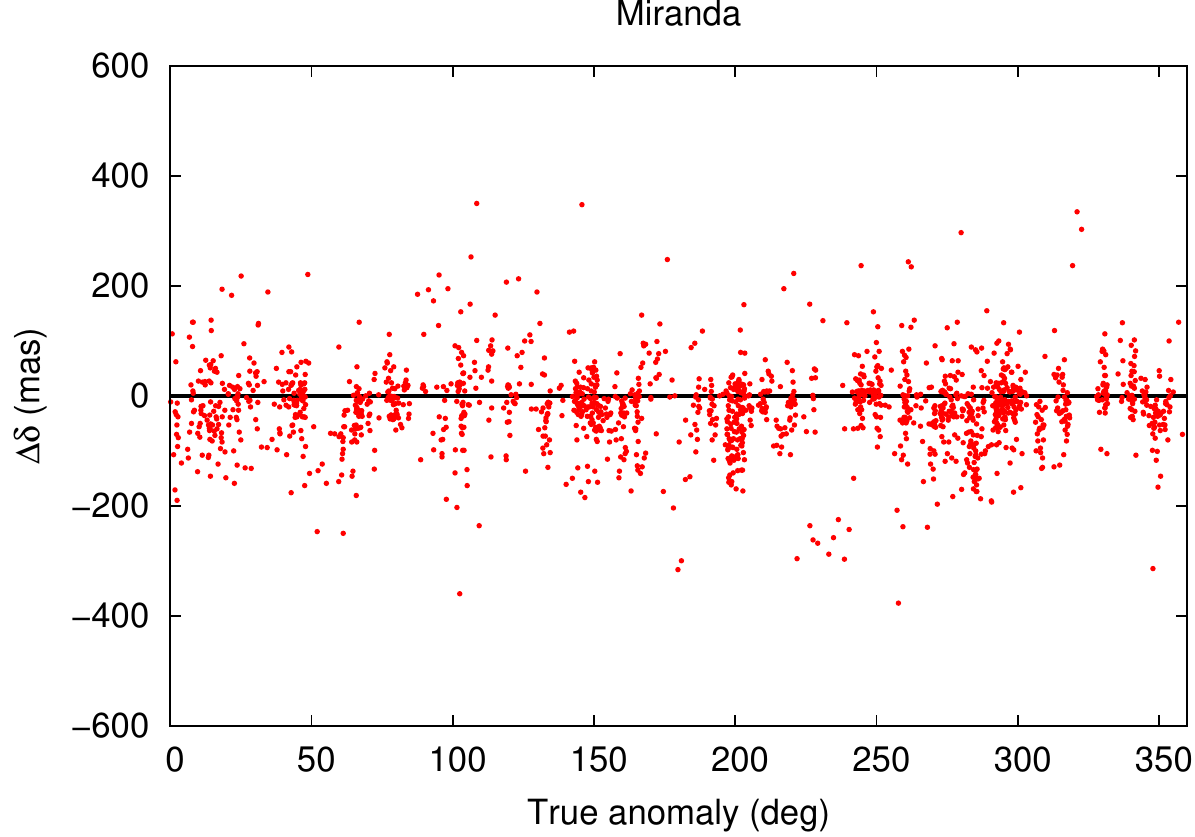}\includegraphics[scale=0.5]{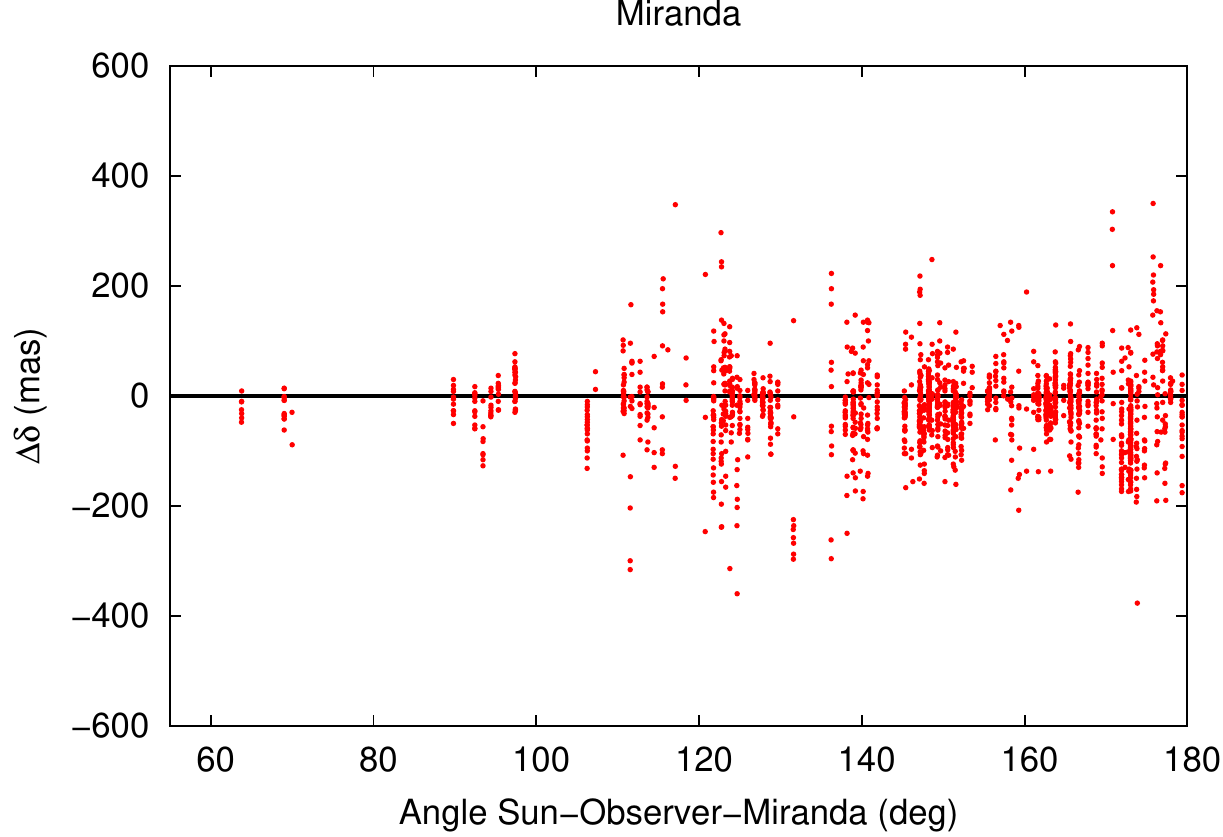}}
\label{fig:miranda}
\end{figure*}

\begin{figure*}
\caption{Same as Fig.~\ref{fig:miranda} for Ariel.}
{\includegraphics[scale=0.5]{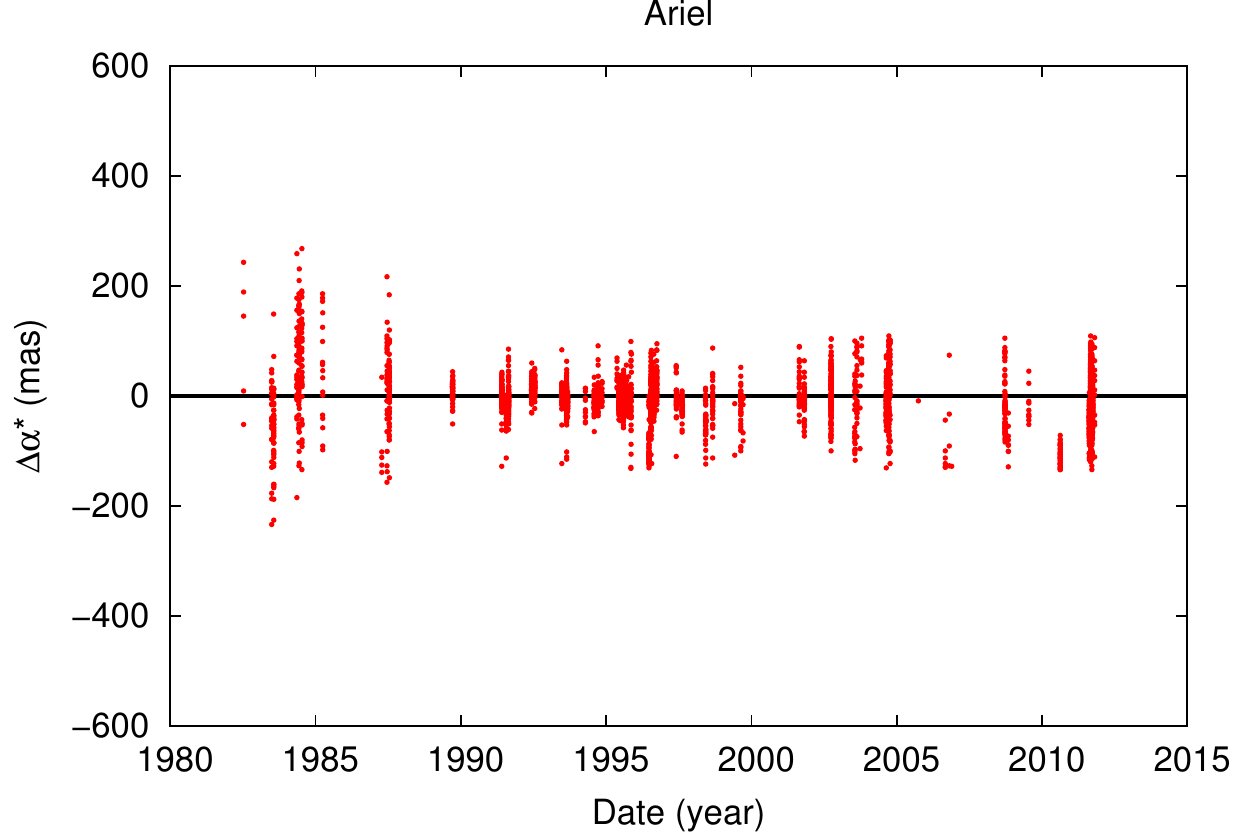}\includegraphics[scale=0.5]{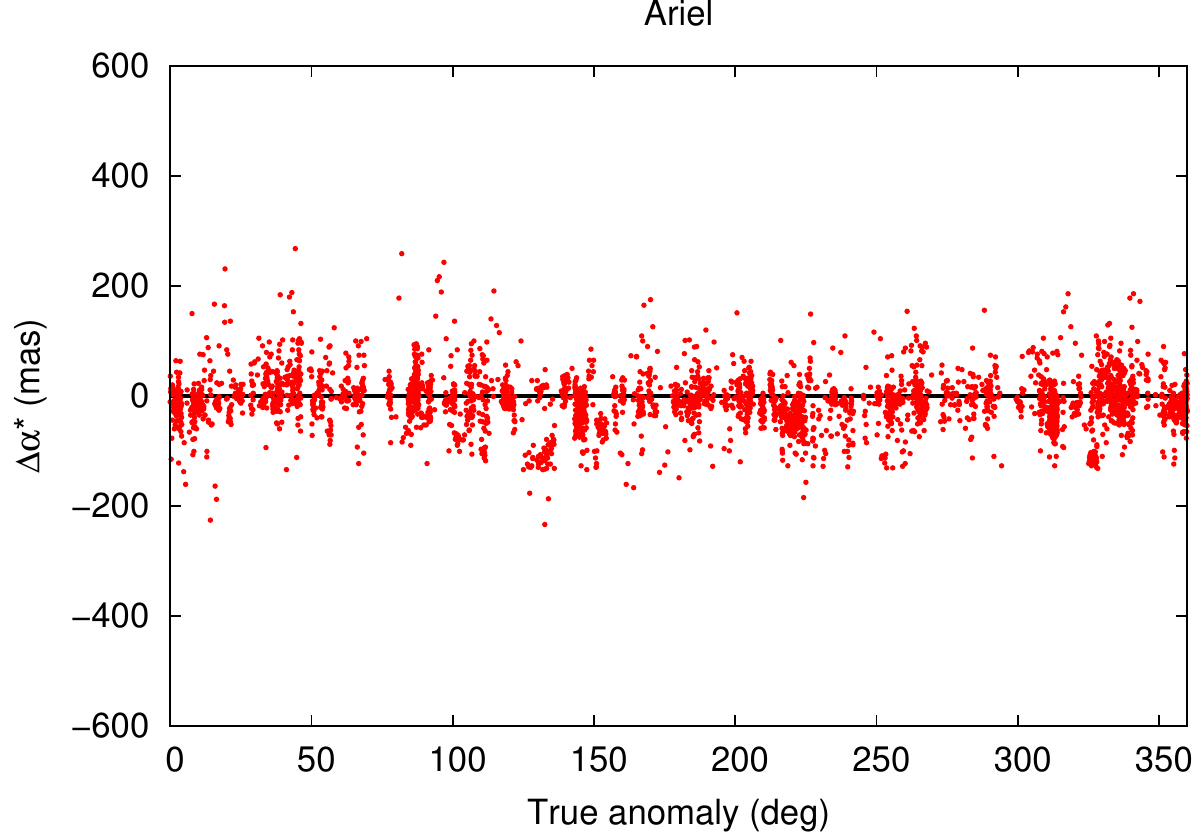}\includegraphics[scale=0.5]{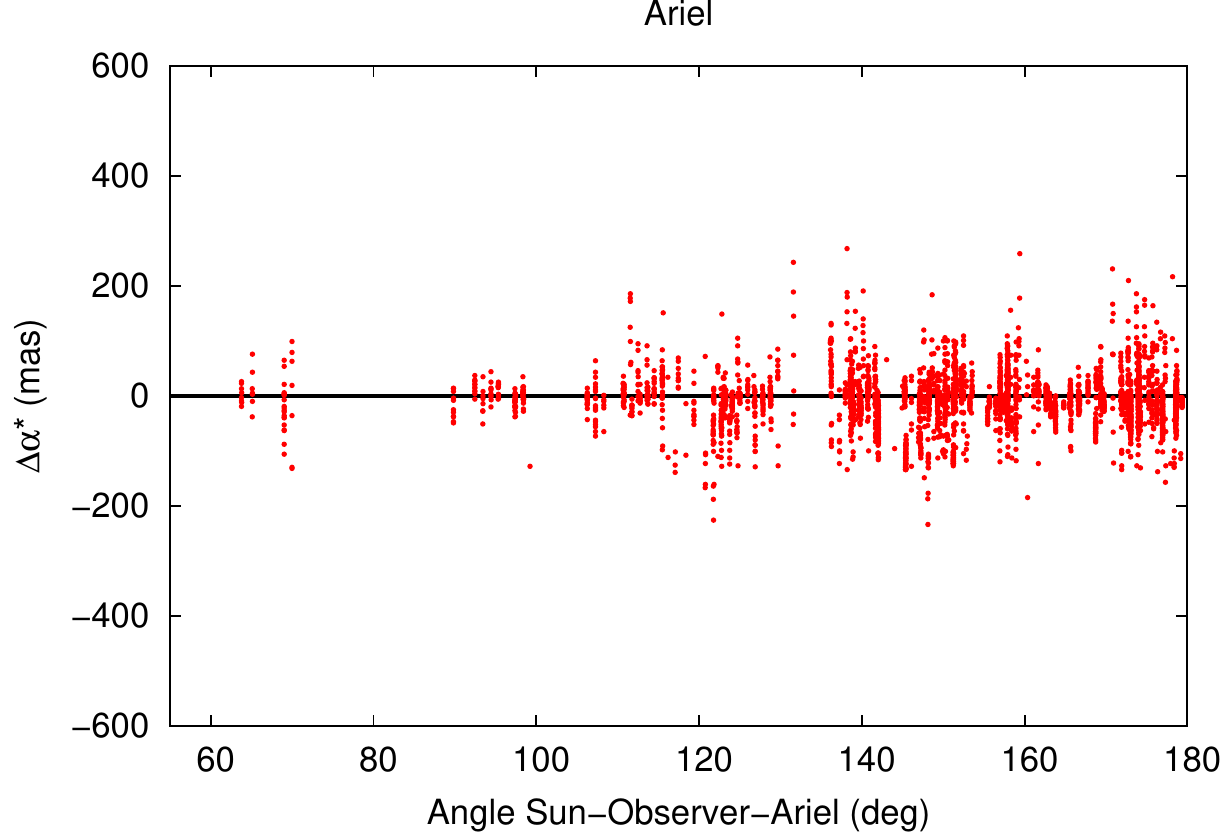}}

{\includegraphics[scale=0.5]{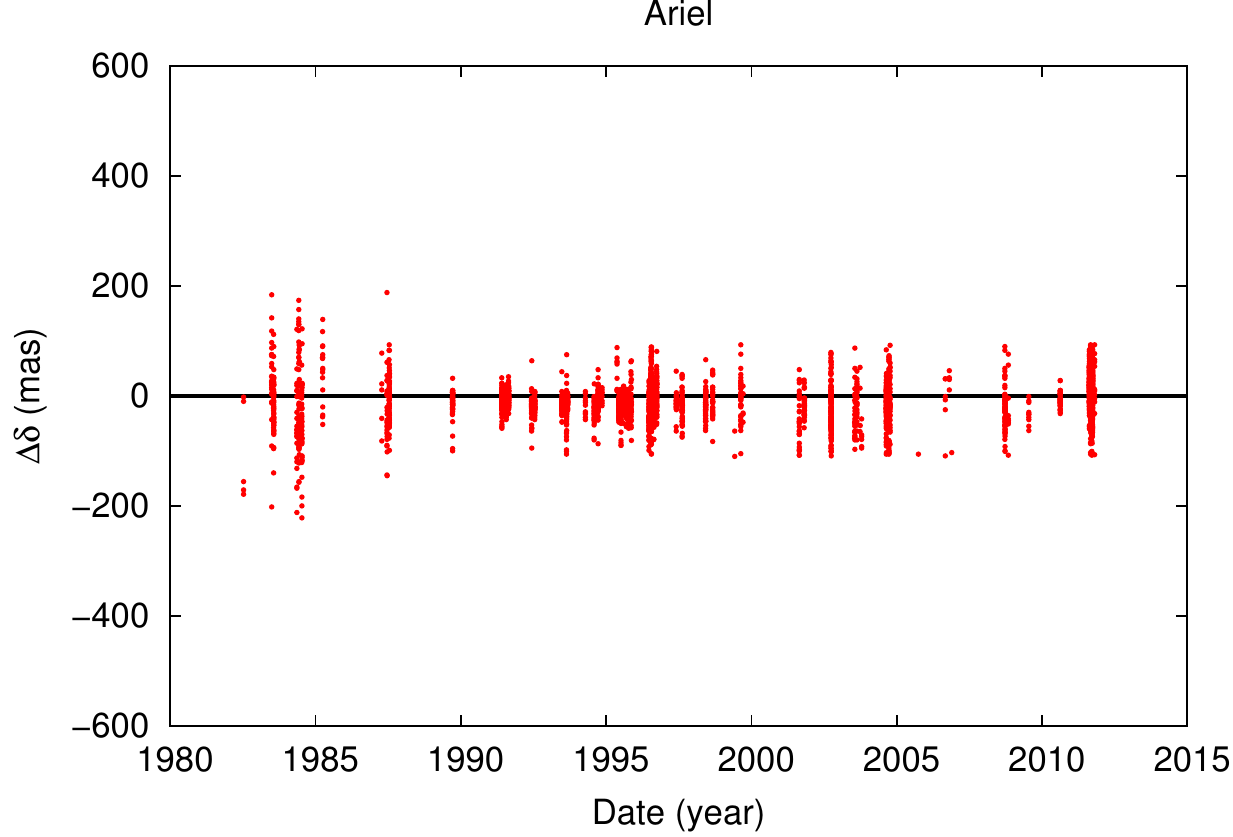}\includegraphics[scale=0.5]{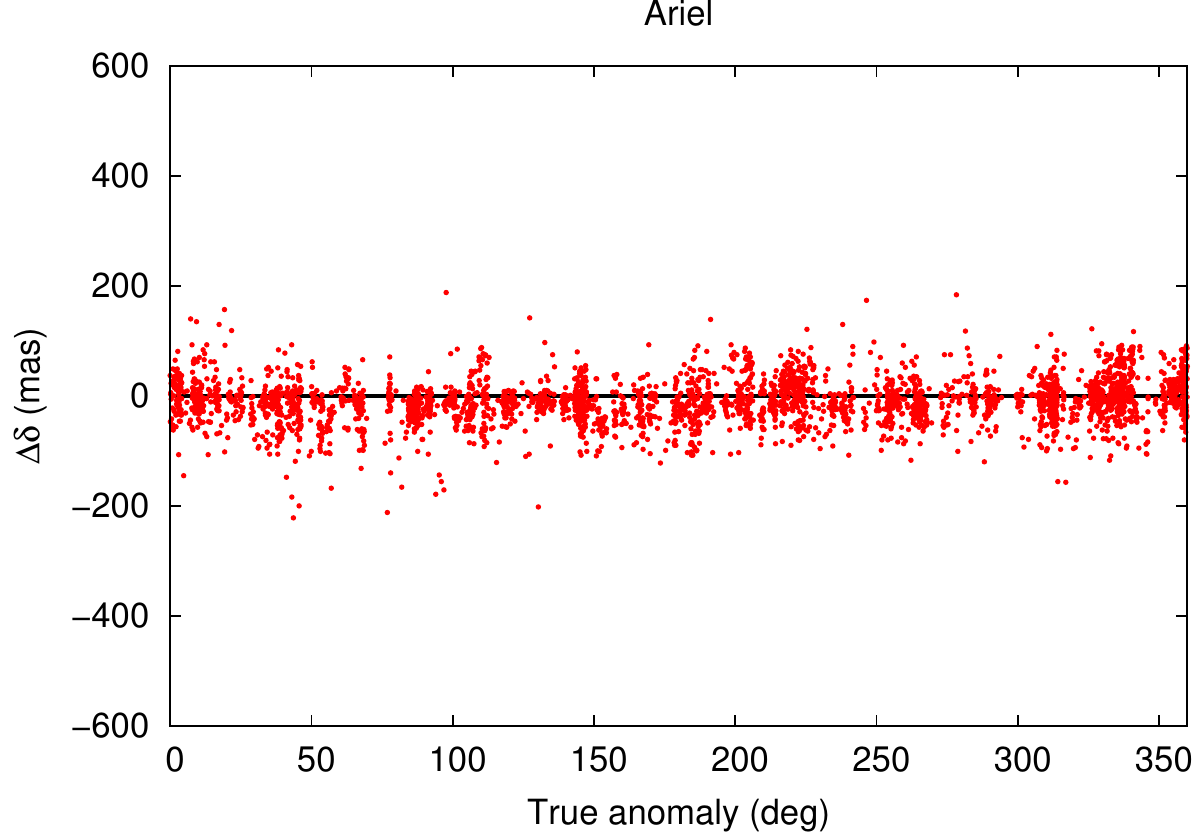}\includegraphics[scale=0.5]{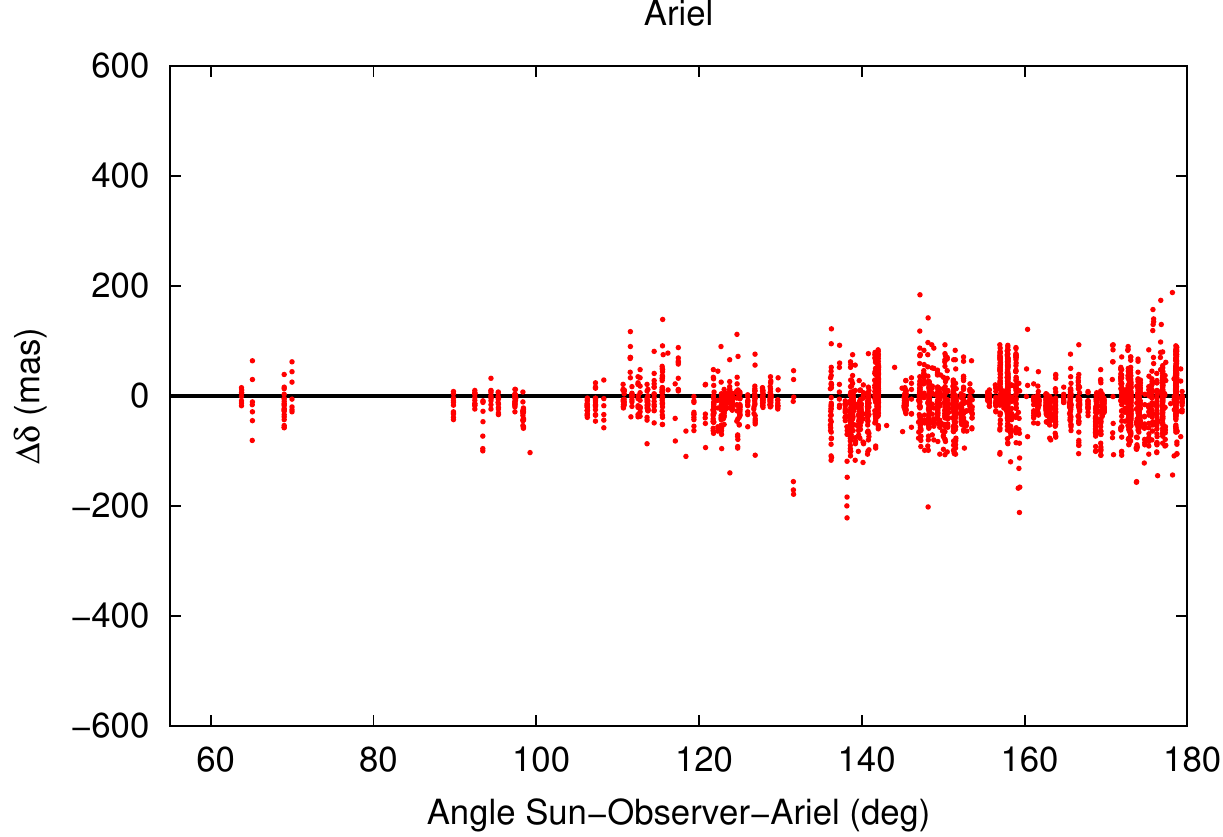}}
\label{fig:ariel}
\end{figure*}

\begin{figure*}
\caption{Same as Fig.~\ref{fig:miranda} for Umbriel.}
{\includegraphics[scale=0.5]{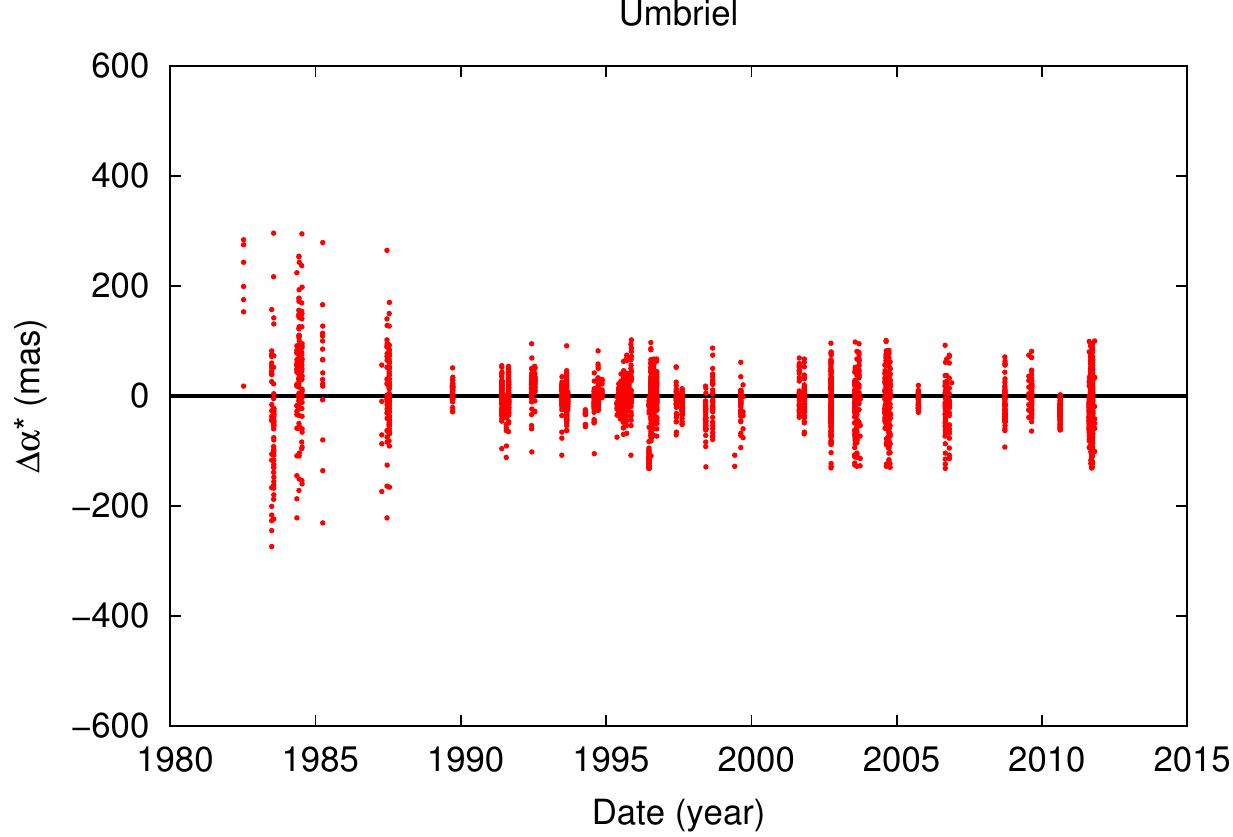}\includegraphics[scale=0.5]{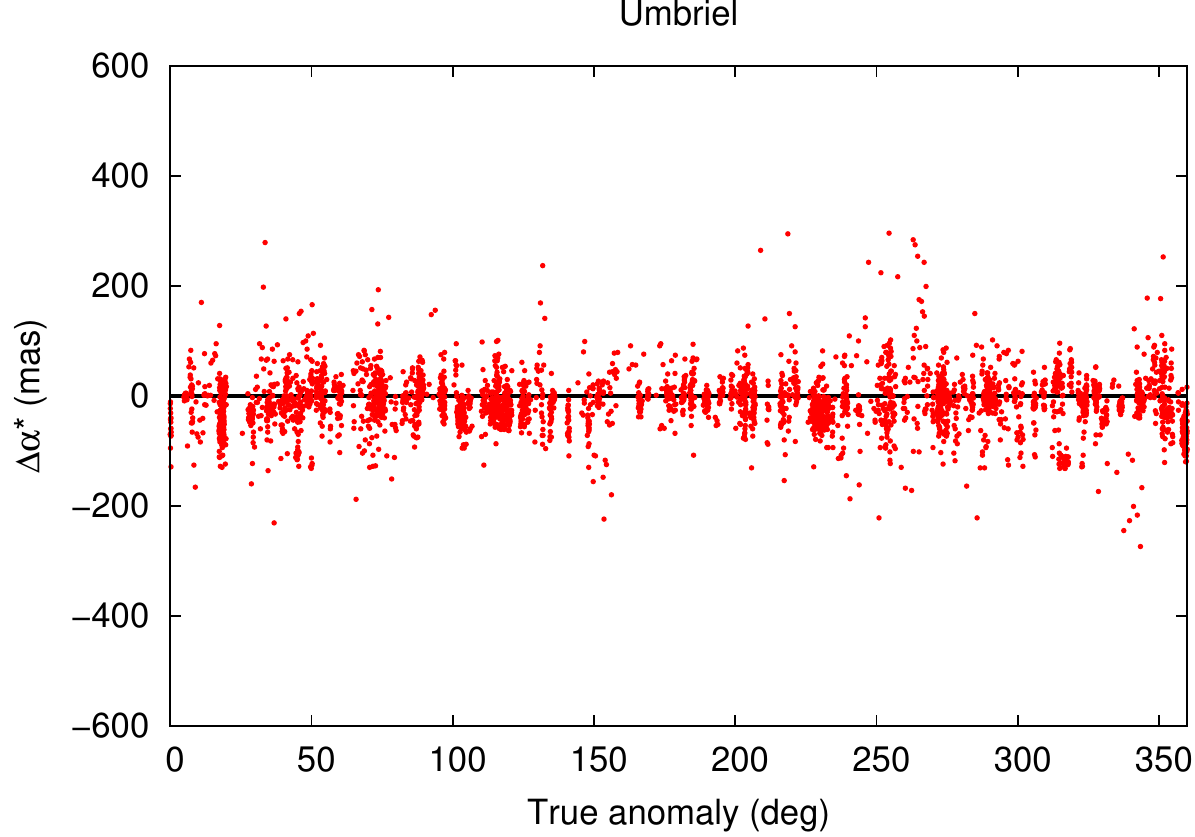}\includegraphics[scale=0.5]{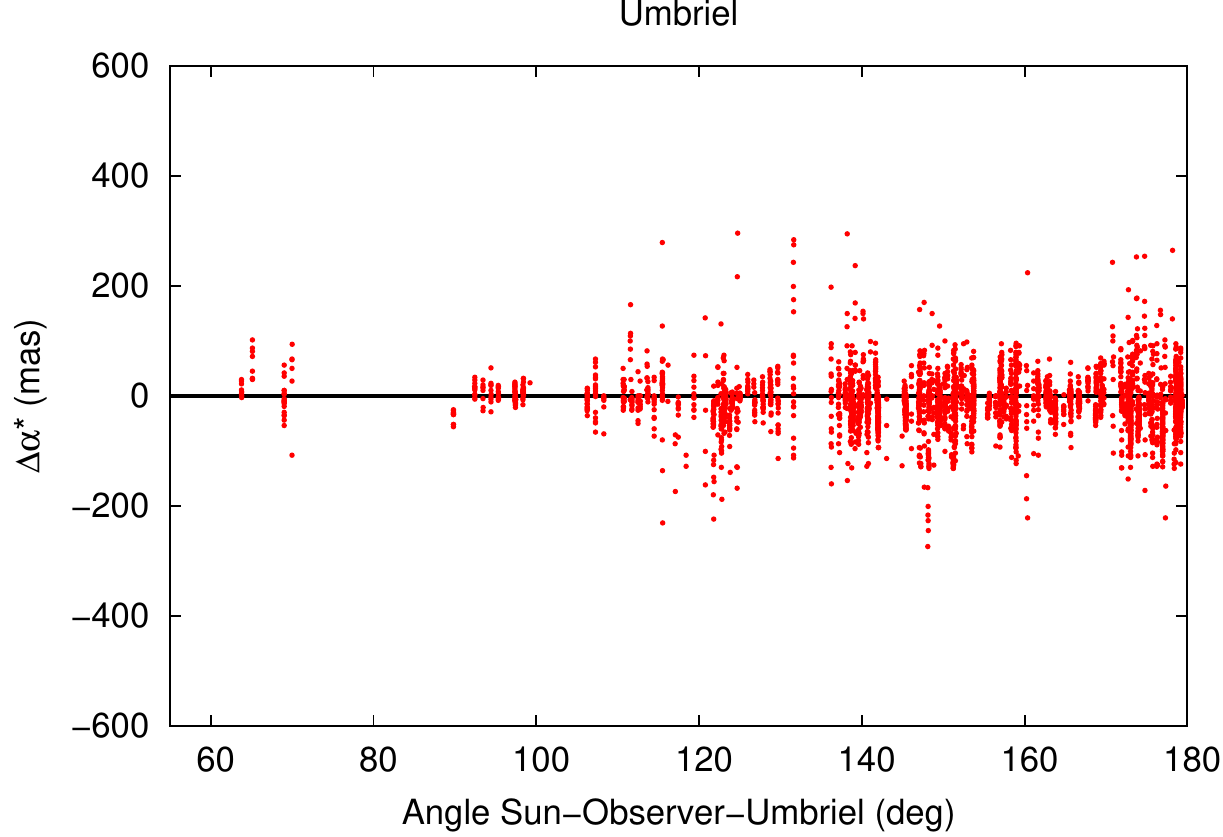}}

{\includegraphics[scale=0.5]{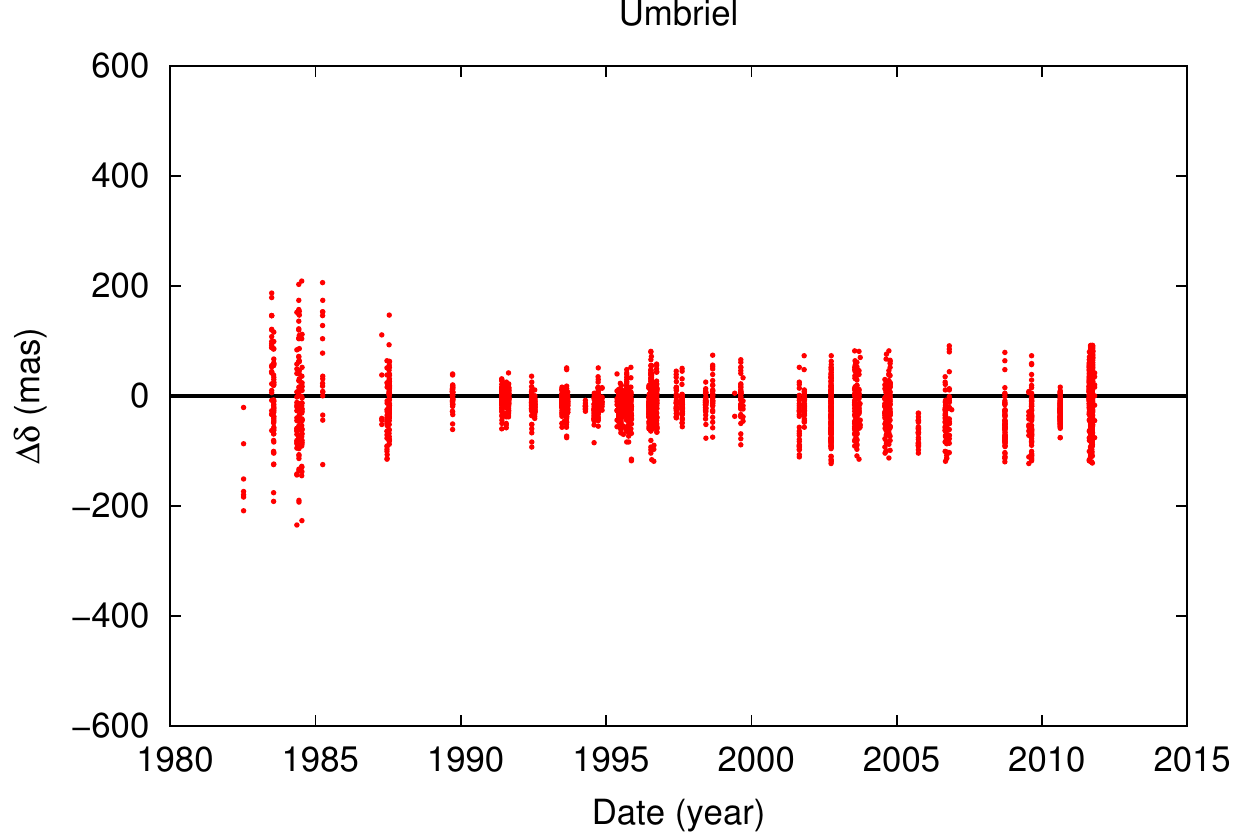}\includegraphics[scale=0.5]{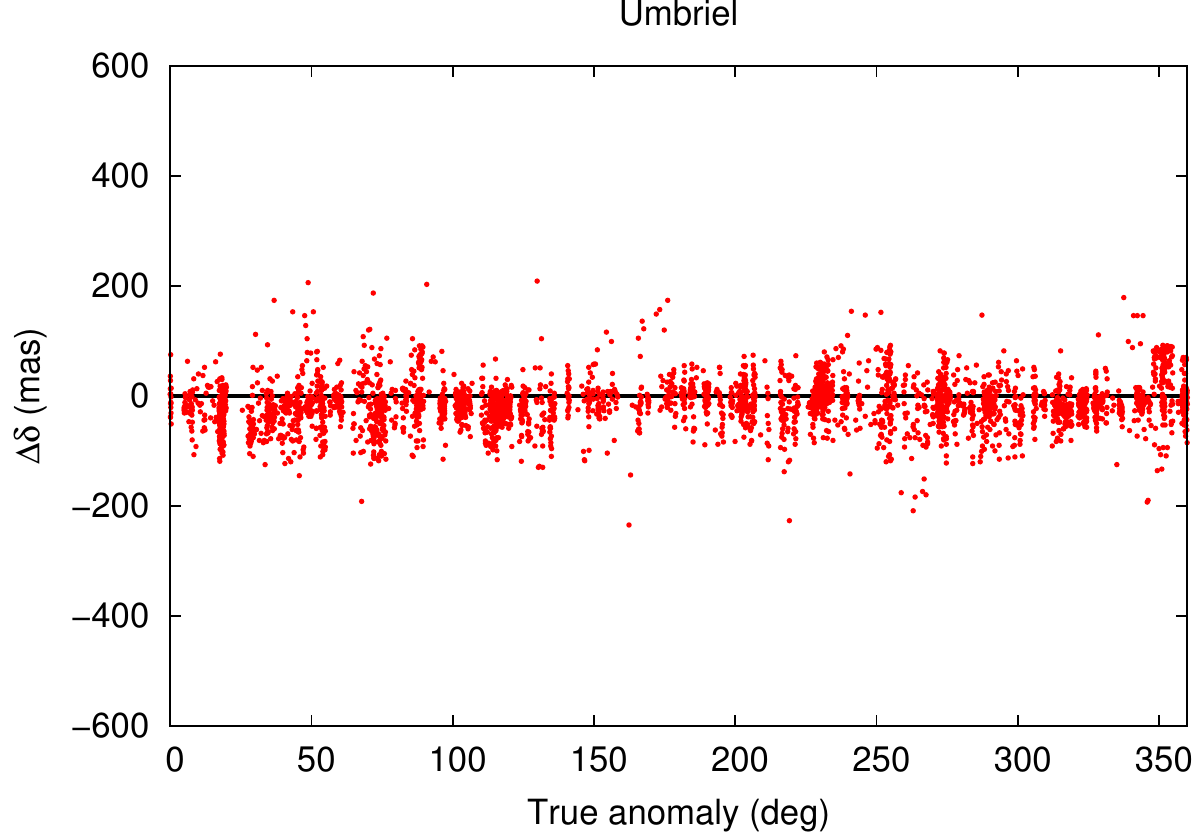}\includegraphics[scale=0.5]{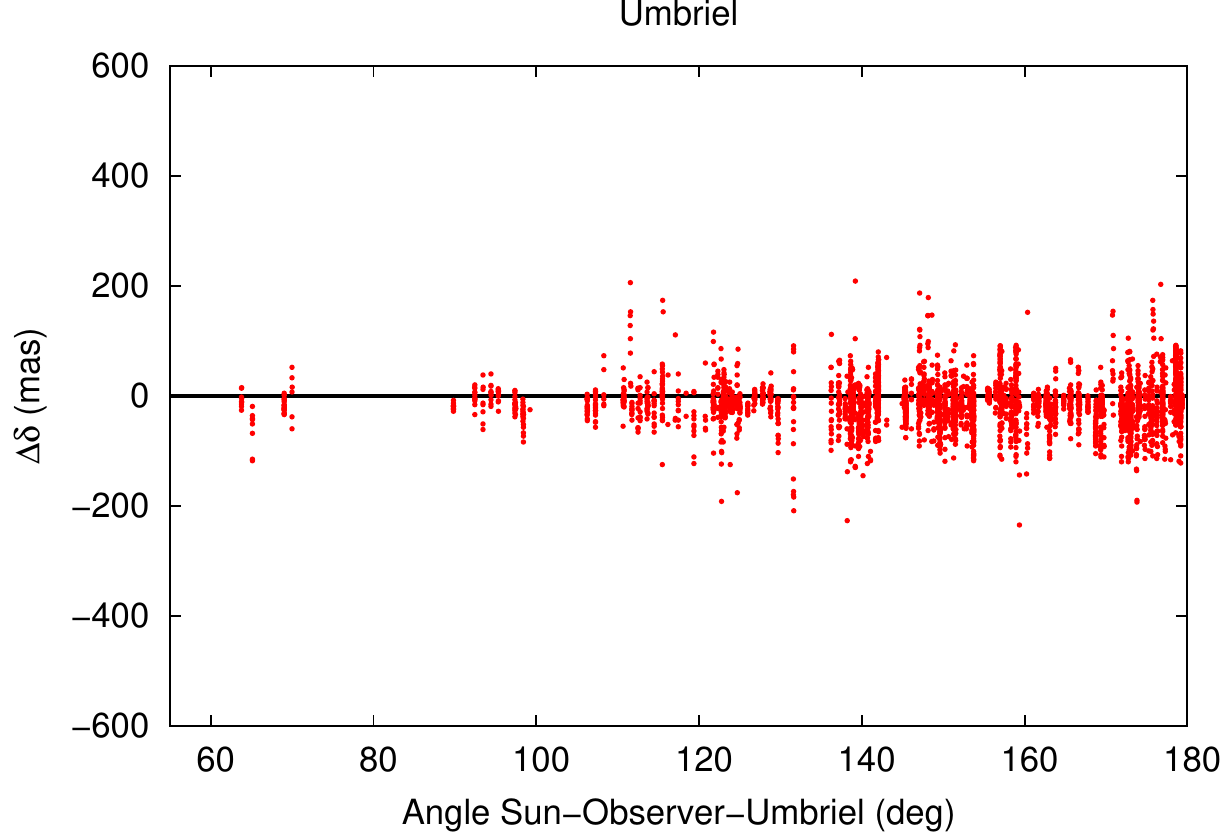}}
\label{fig:umbriel}
\end{figure*}

\begin{figure*}
\caption{Same as Fig.~\ref{fig:miranda} for Titania.}
{\includegraphics[scale=0.5]{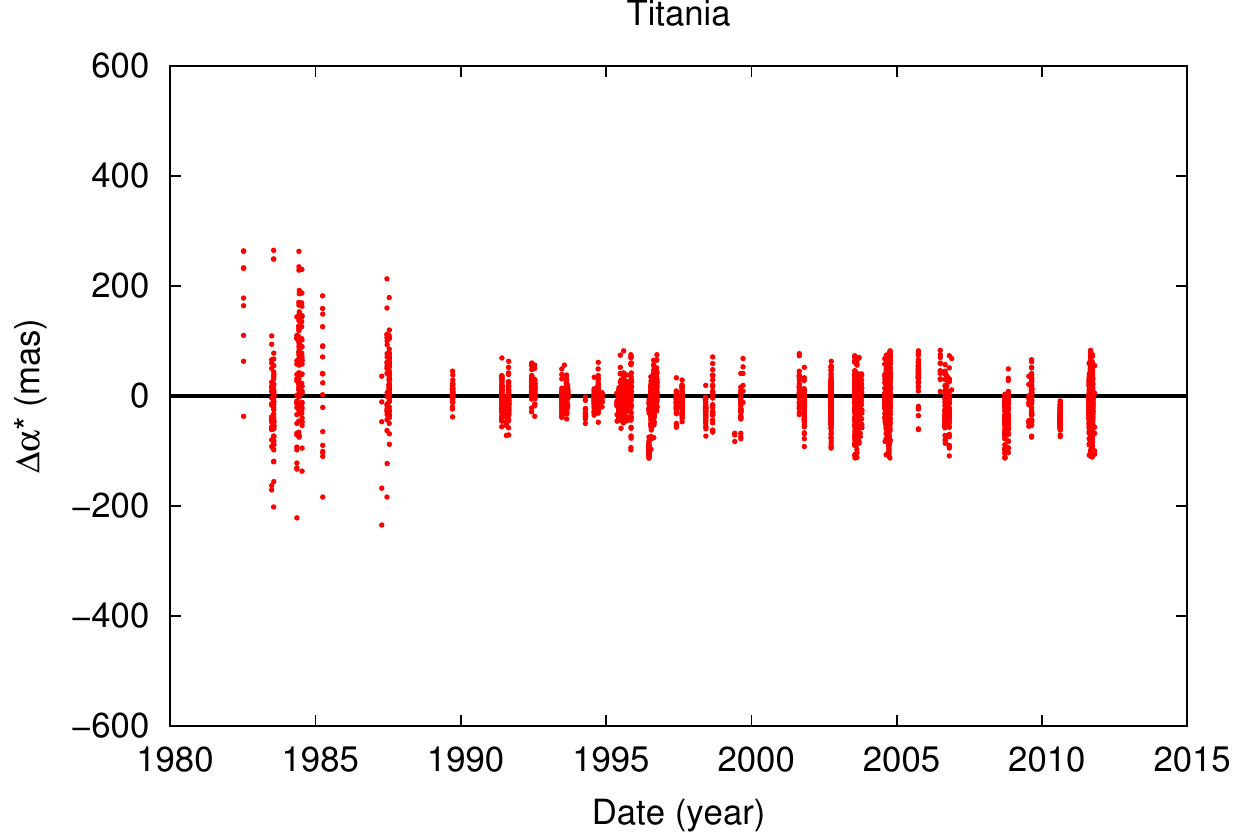}\includegraphics[scale=0.5]{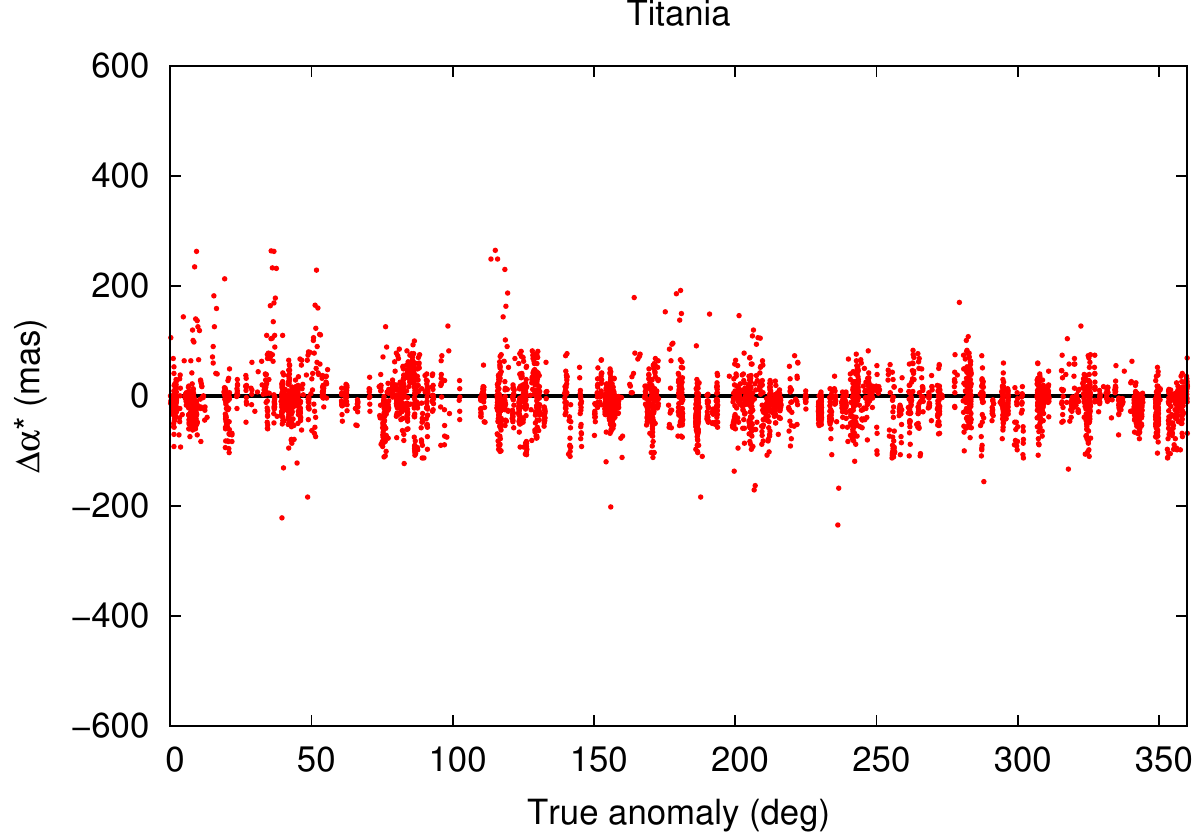}\includegraphics[scale=0.5]{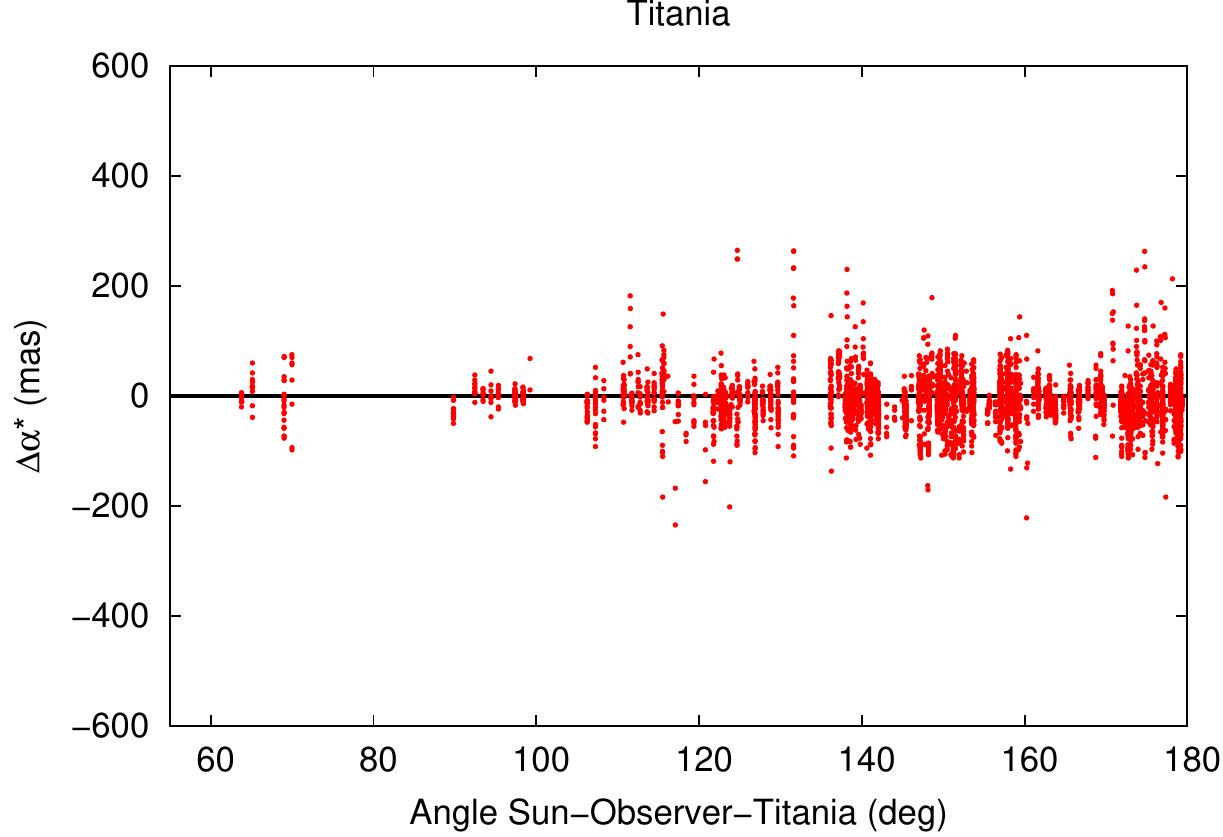}}

{\includegraphics[scale=0.5]{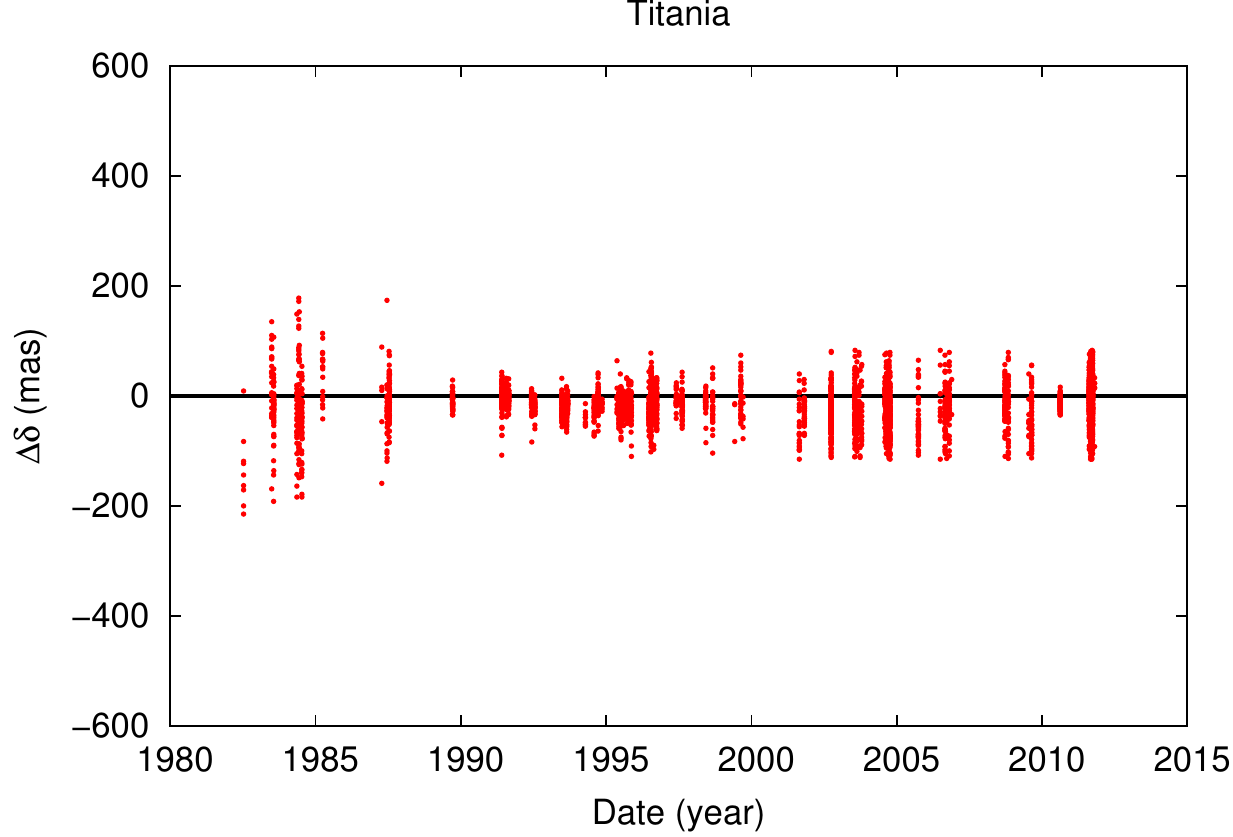}\includegraphics[scale=0.5]{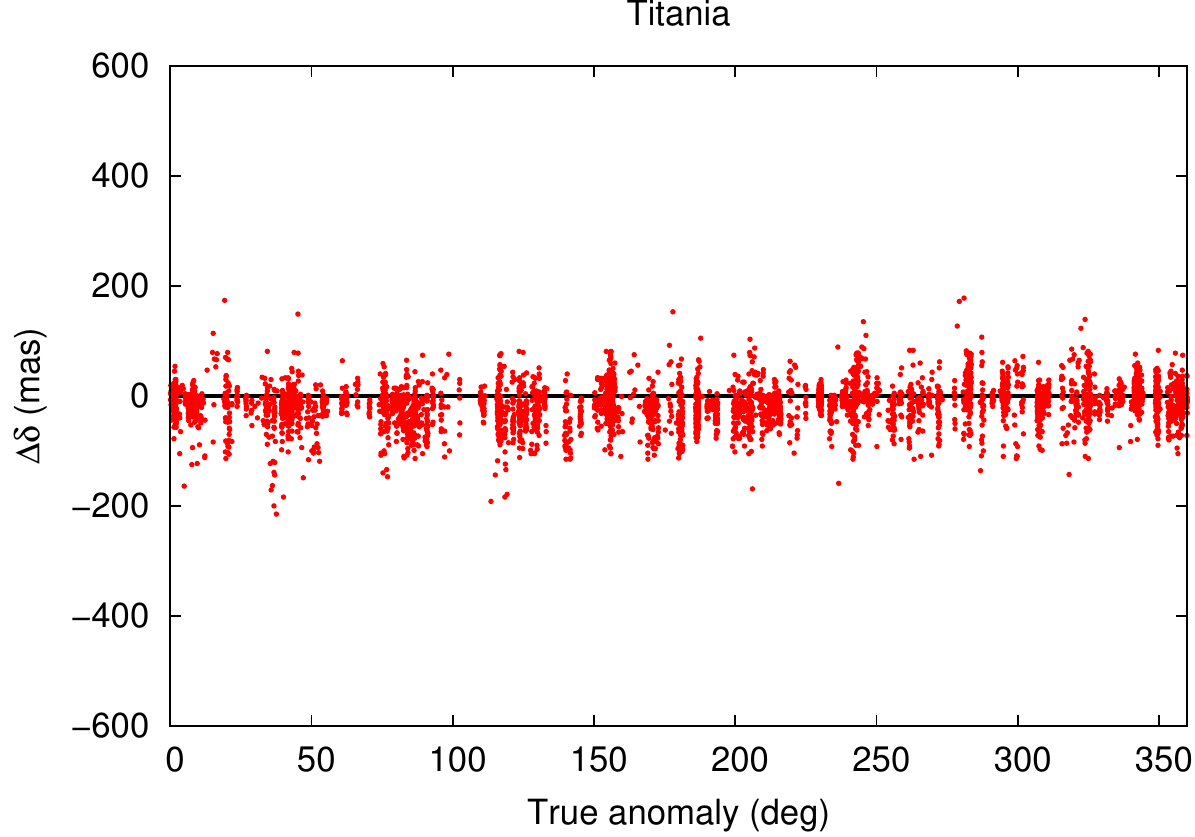}\includegraphics[scale=0.5]{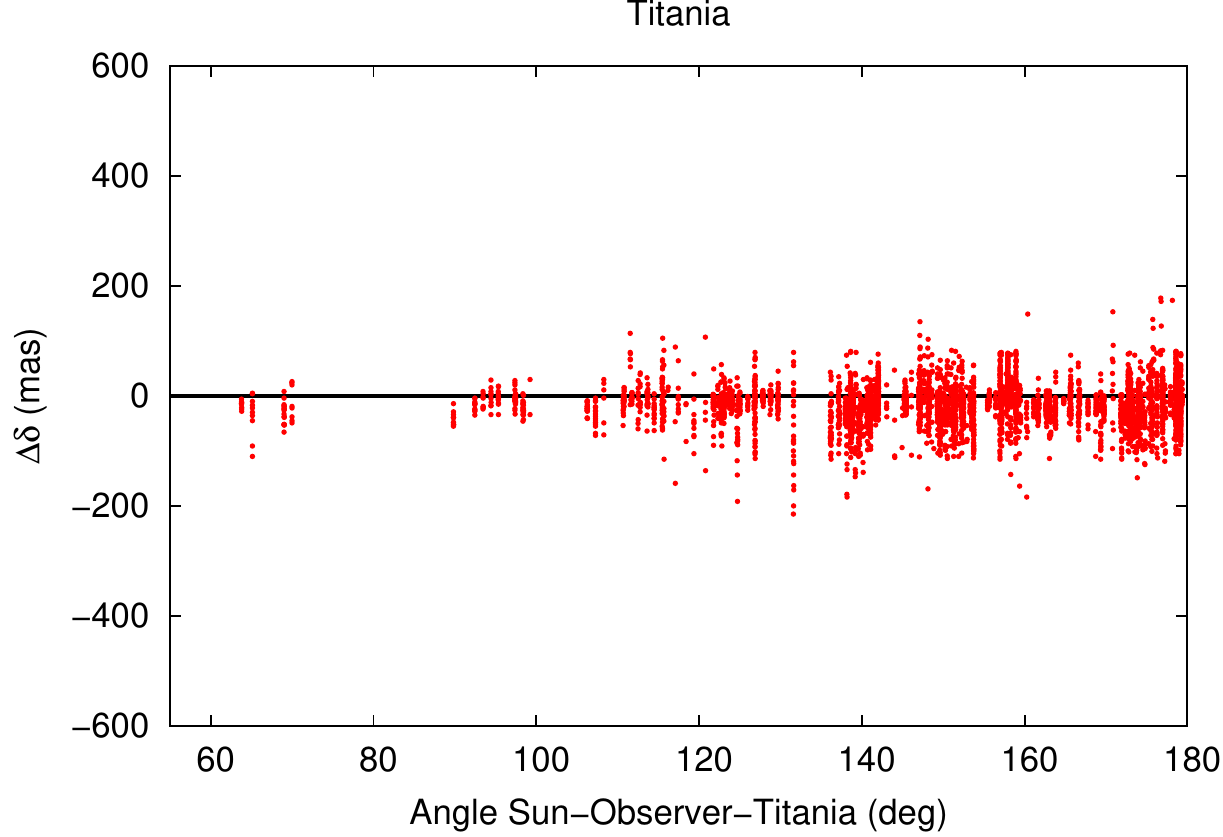}}
\label{fig:titania}
\end{figure*}

\begin{figure*}
\caption{Same as Fig.~\ref{fig:miranda} for Oberon.}
{\includegraphics[scale=0.5]{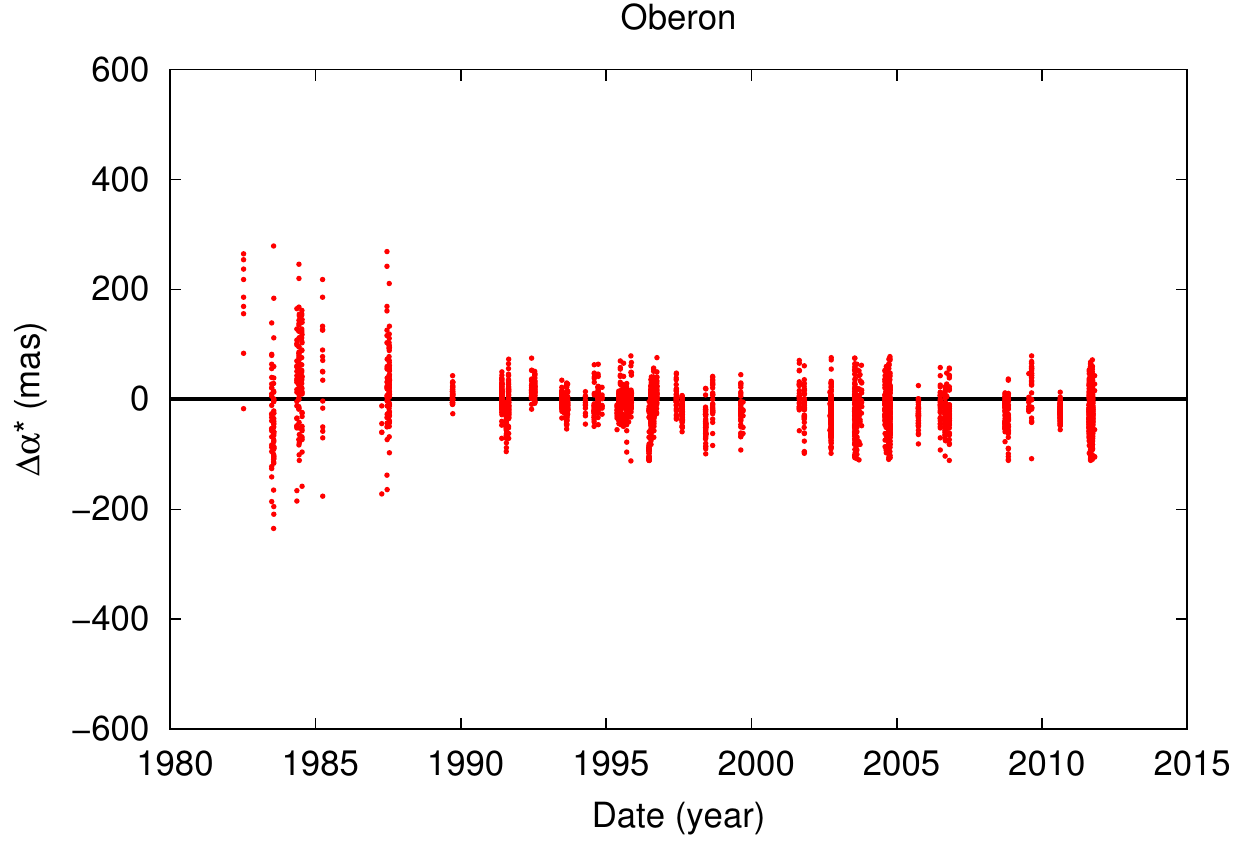}\includegraphics[scale=0.5]{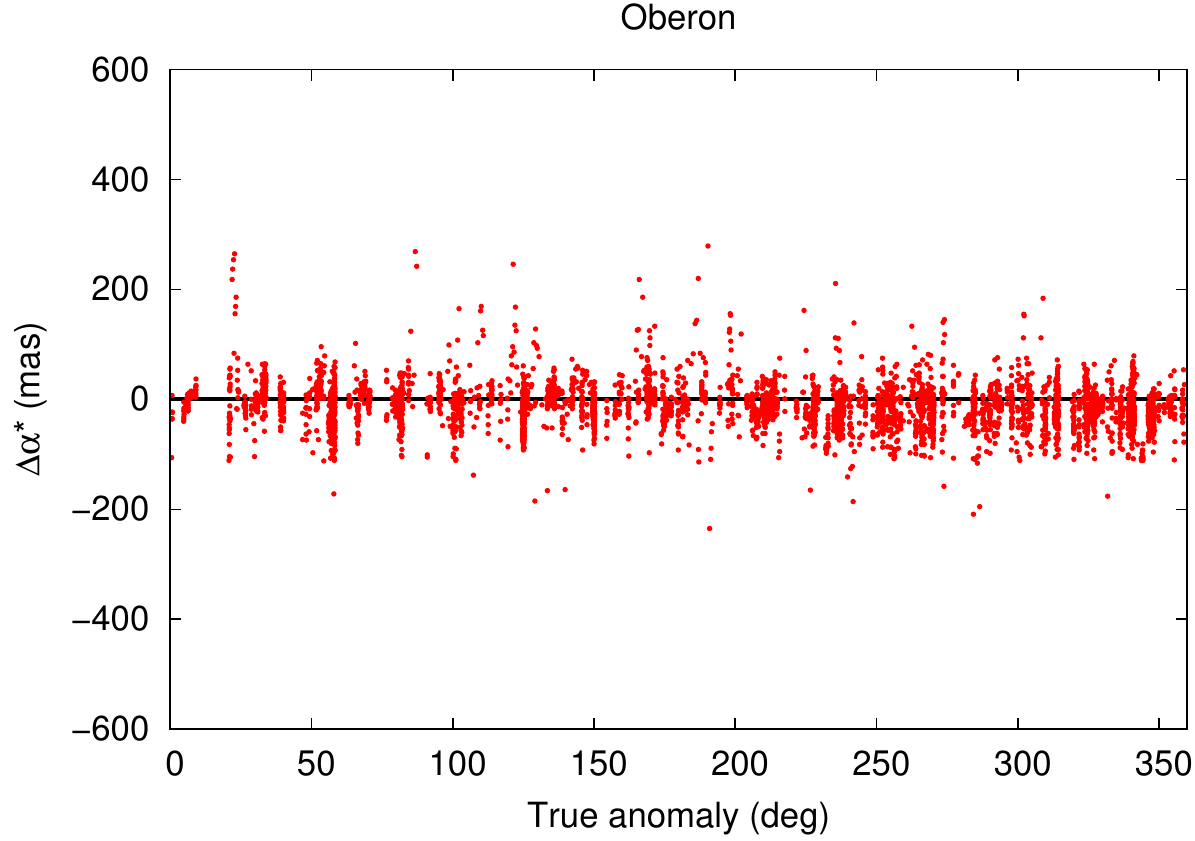}\includegraphics[scale=0.5]{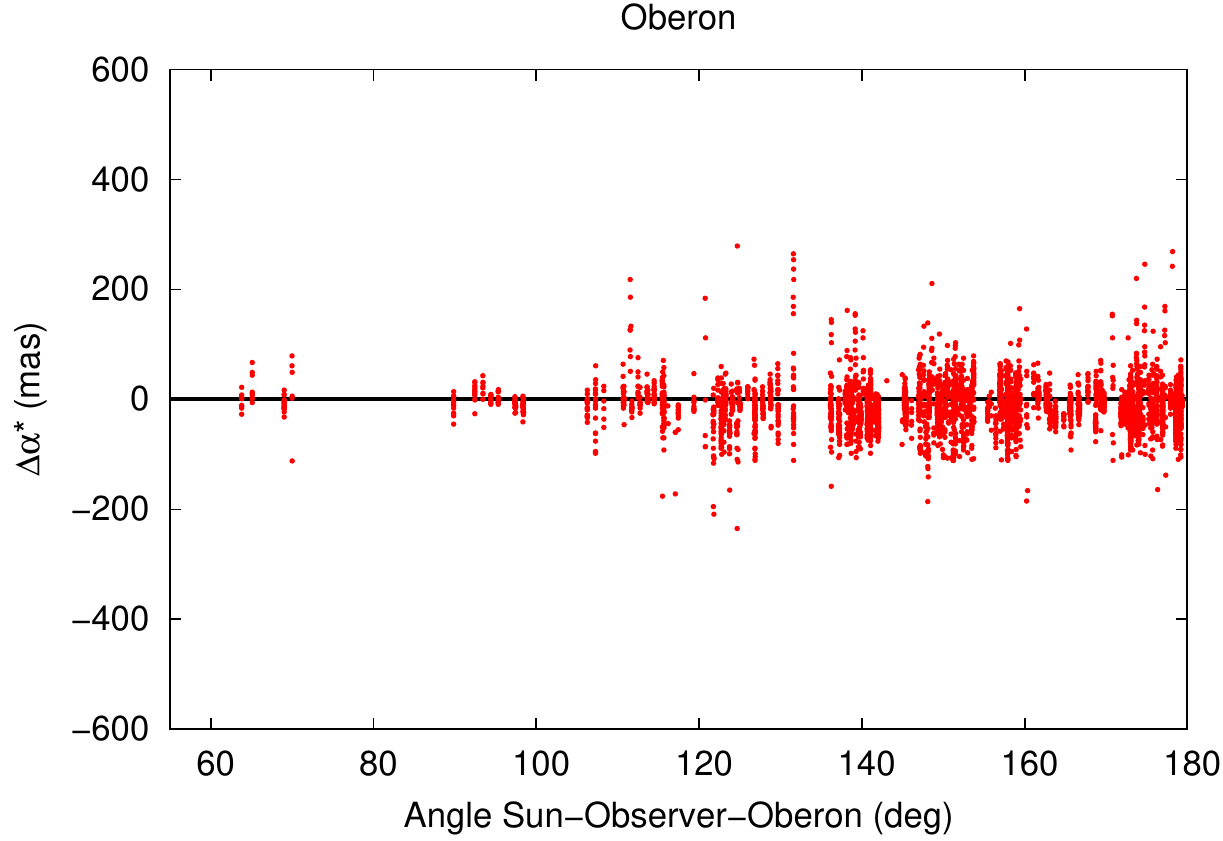}}

{\includegraphics[scale=0.5]{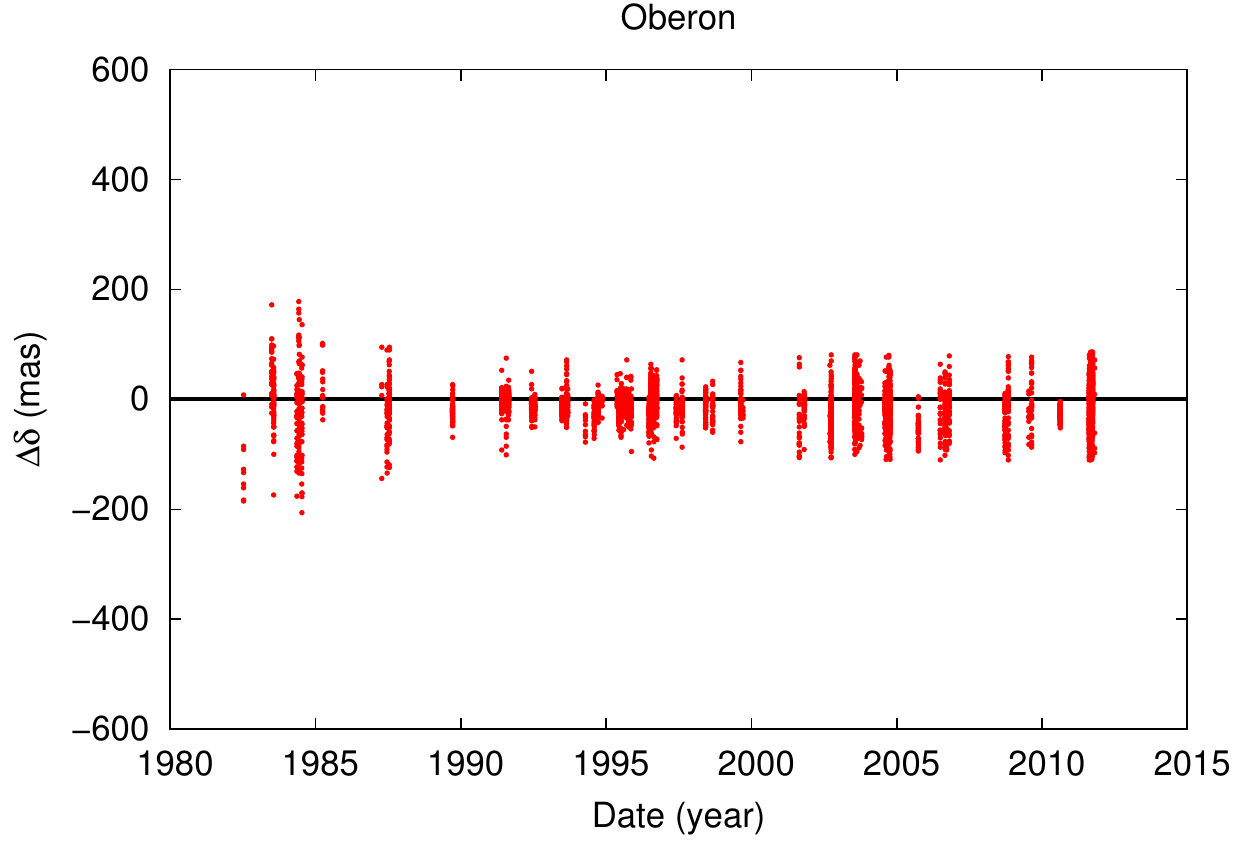}\includegraphics[scale=0.5]{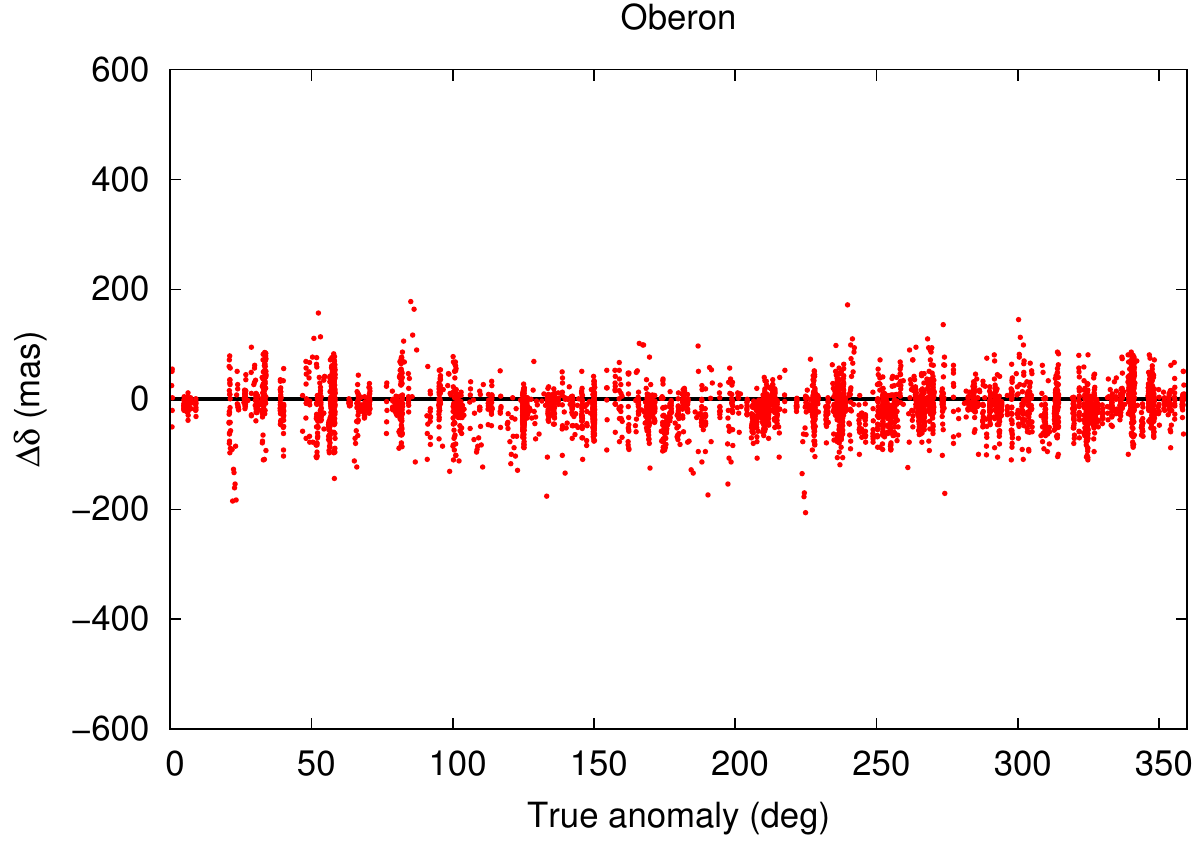}\includegraphics[scale=0.5]{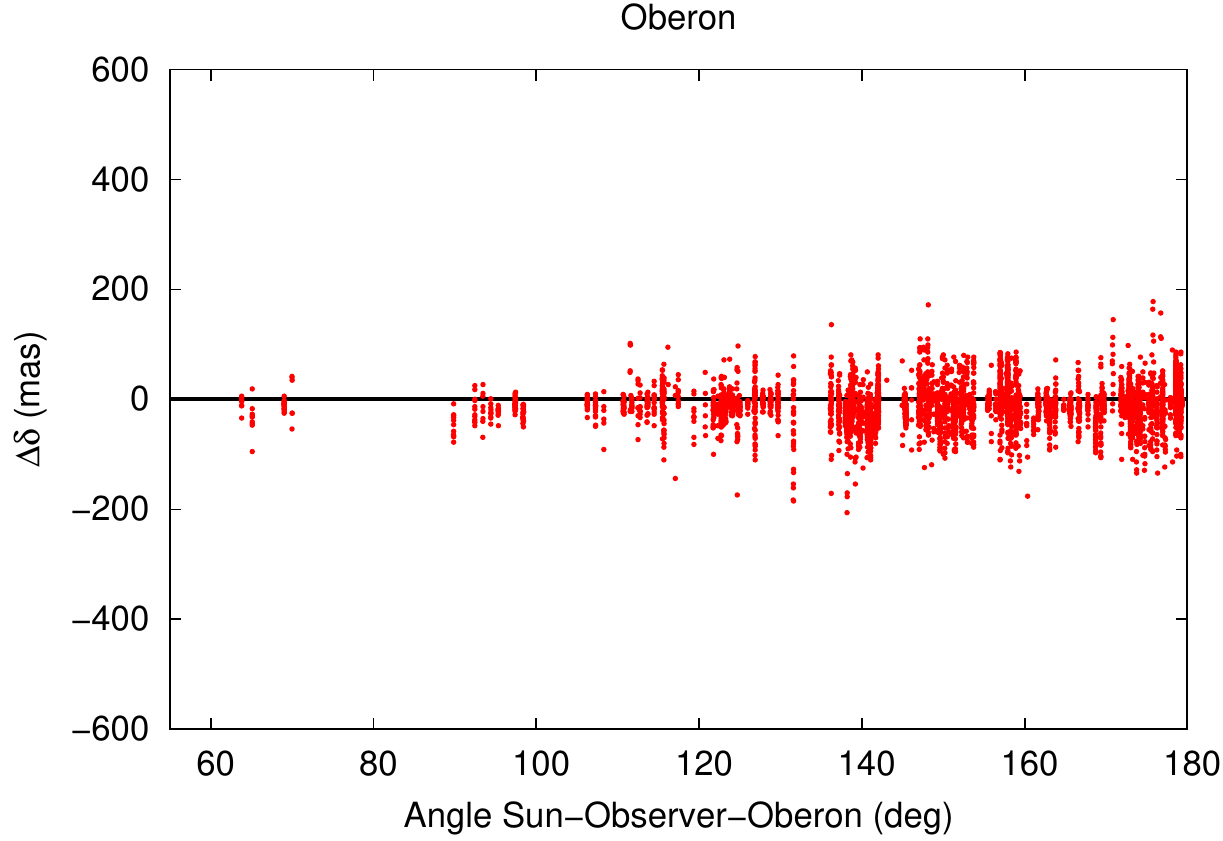}}
\label{fig:oberon}
\end{figure*}

\begin{figure*}
\caption{Differences in right ascension (left panel) and declination (middle panel) in the sense observations minus ephemerides (DE435+ura111) for Miranda, as a function of the angular distance from Uranus. Right panel: Angular distance of Miranda from Uranus as a function of time. The "*" means multiplication by the cosine of the declination.}
{\includegraphics[scale=0.5]{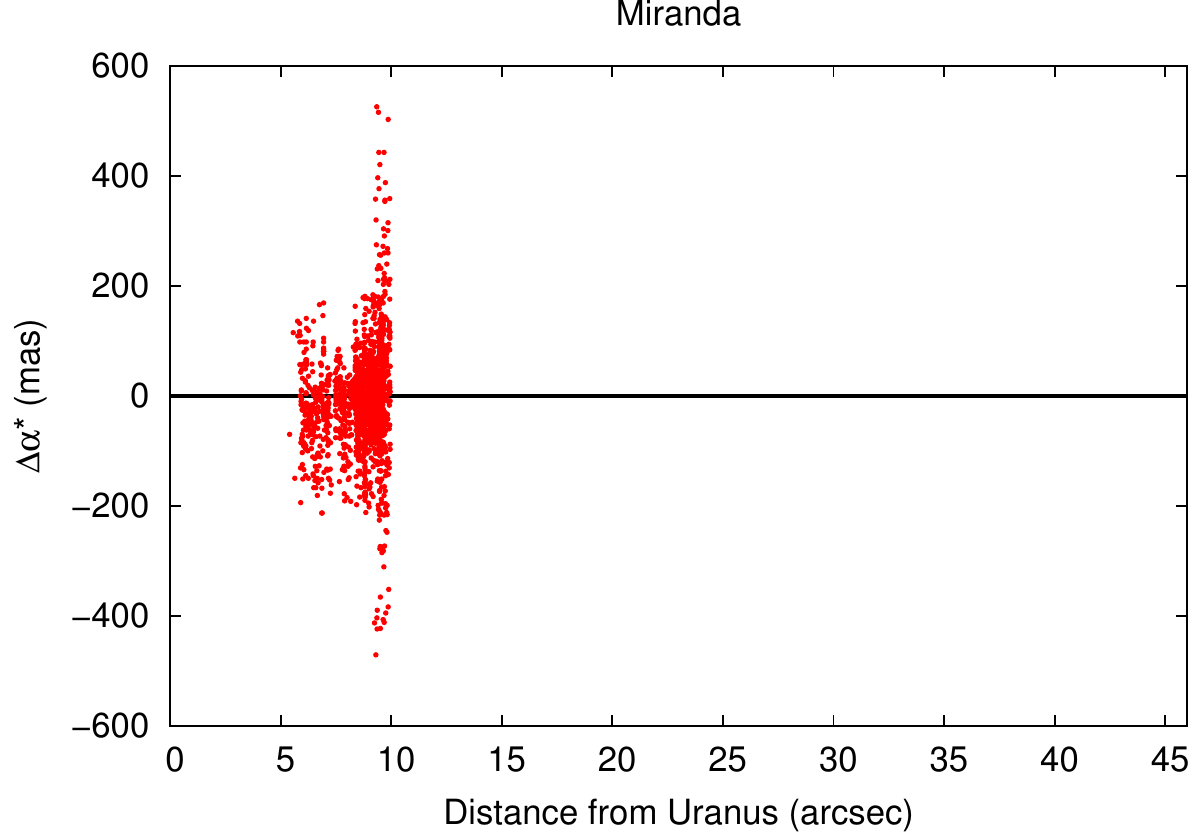}\includegraphics[scale=0.5]{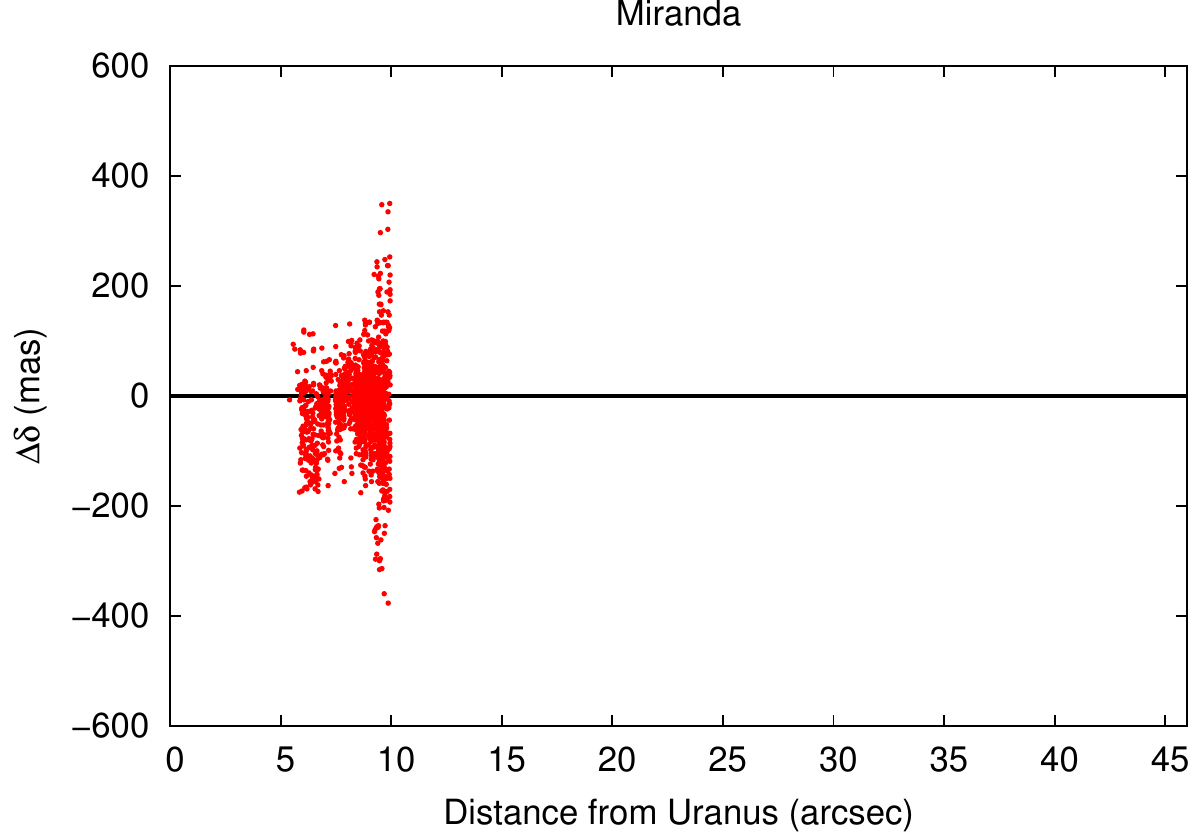}\includegraphics[scale=0.5]{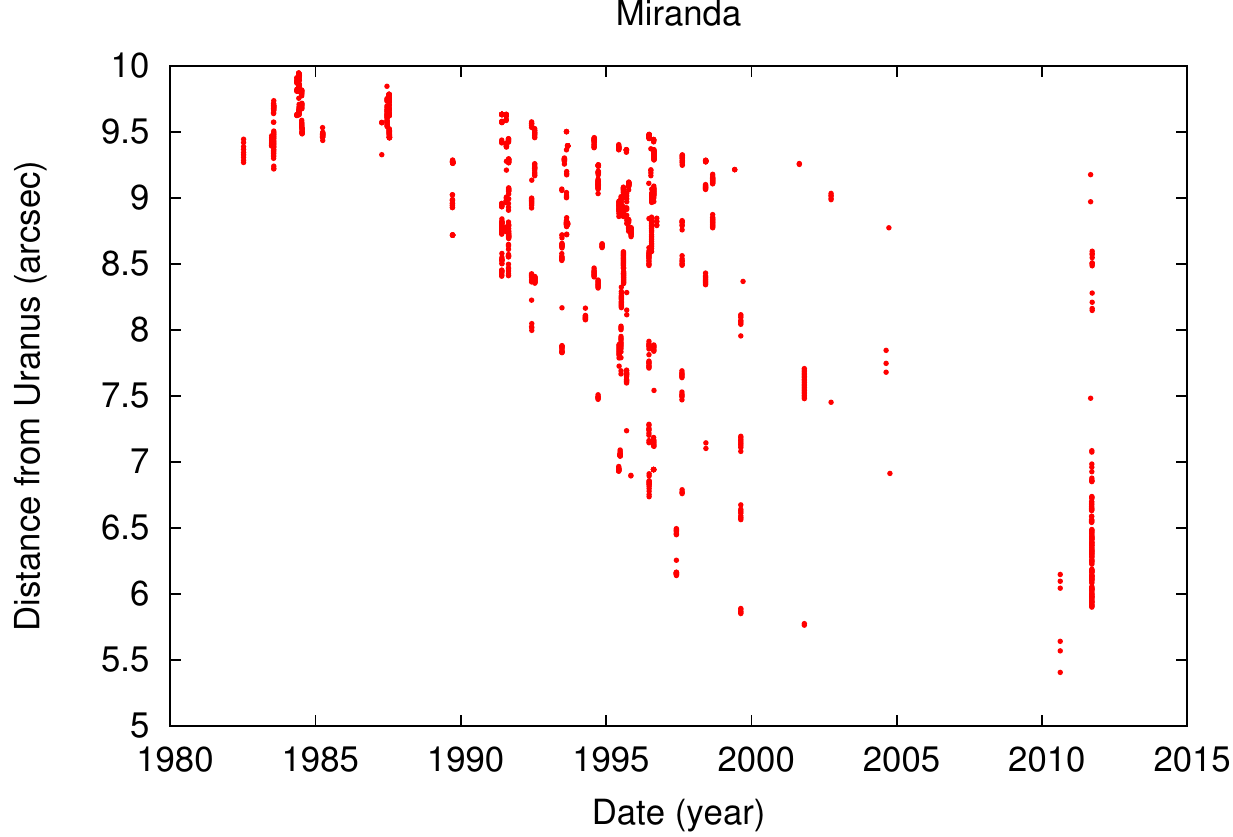}}
\label{fig:miranda_dist}
\end{figure*}

\begin{figure*}
\caption{Same as Fig.~\ref{fig:miranda_dist} for Ariel.}
{\includegraphics[scale=0.5]{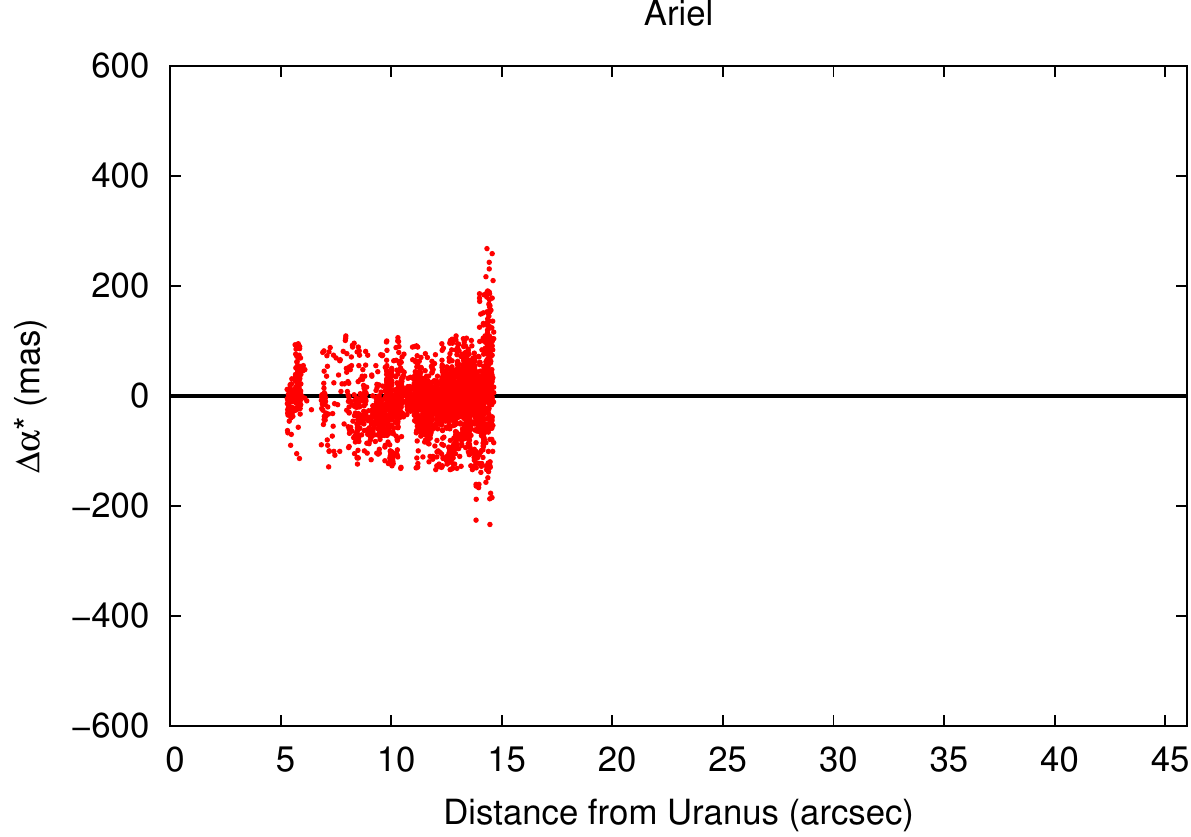}\includegraphics[scale=0.5]{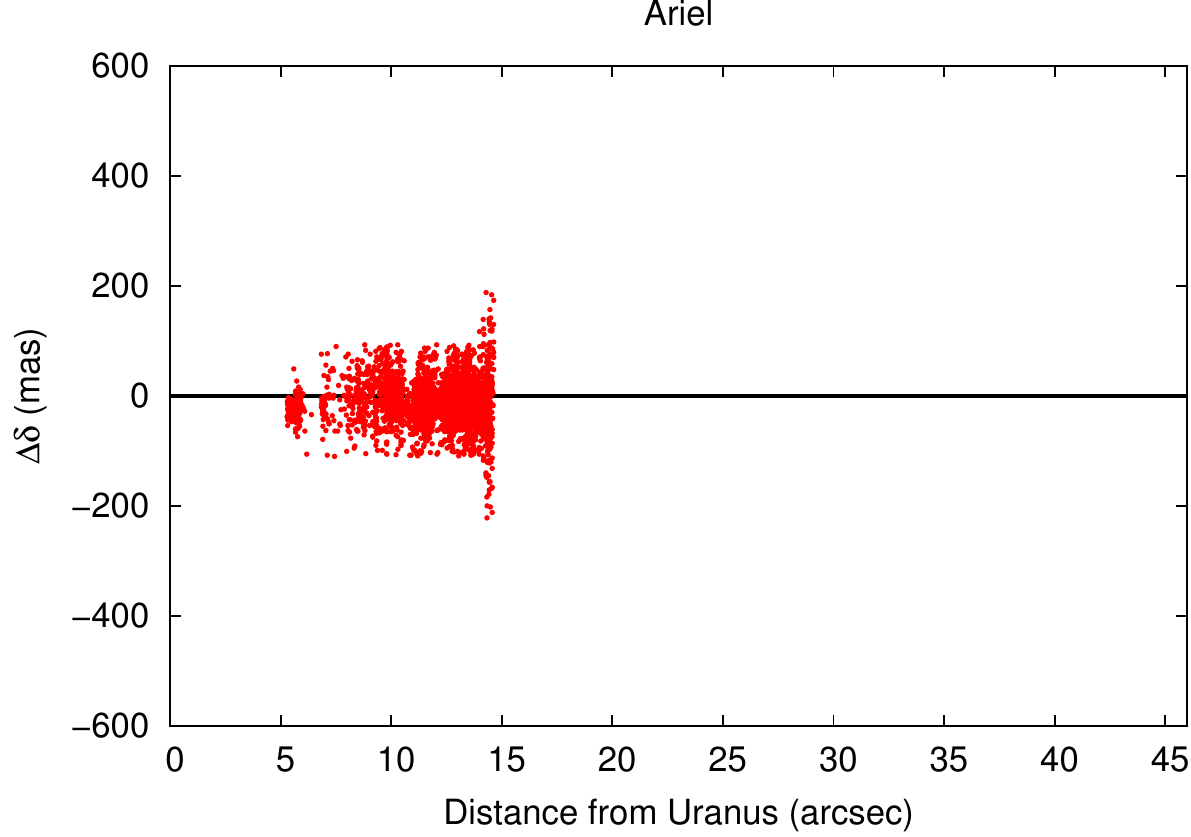}\includegraphics[scale=0.5]{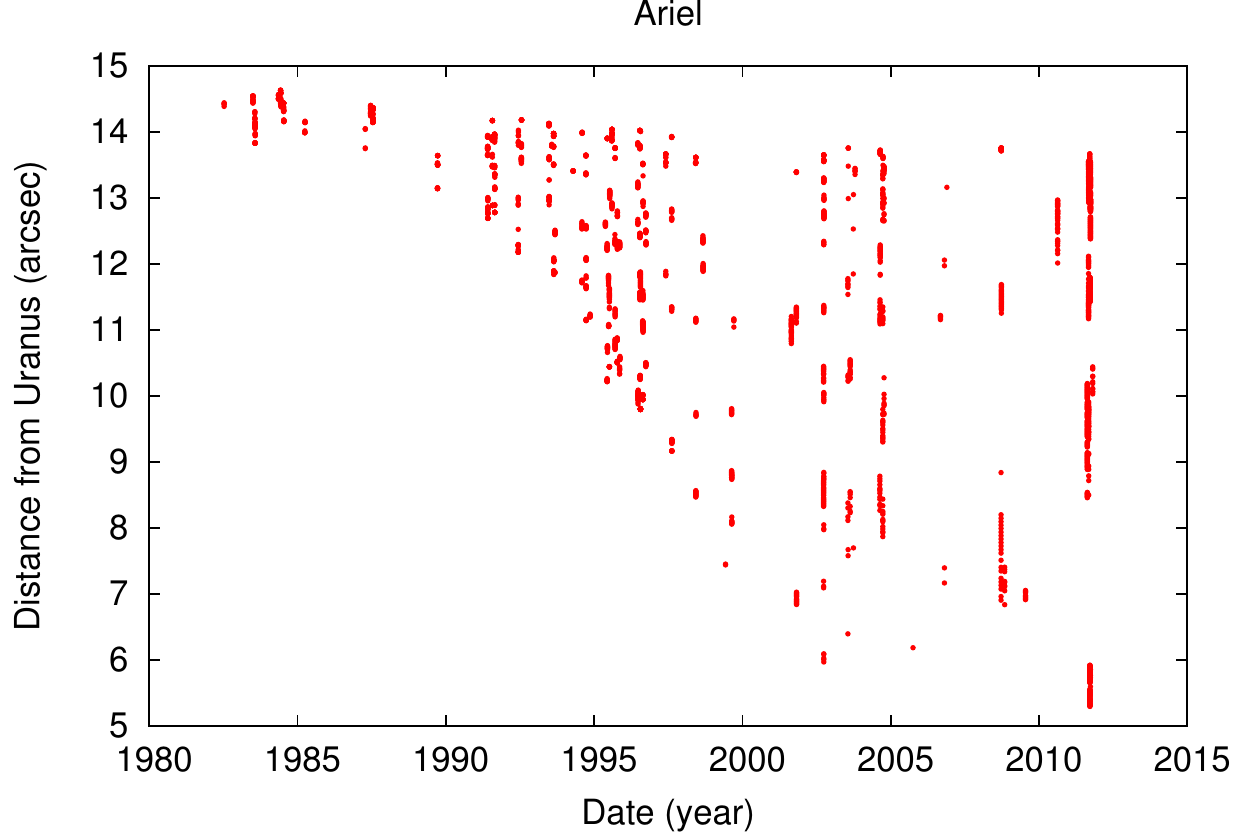}}
\label{fig:ariel_dist}
\end{figure*}

\begin{figure*}
\caption{Same as Fig.~\ref{fig:miranda_dist} for Umbriel.}
{\includegraphics[scale=0.5]{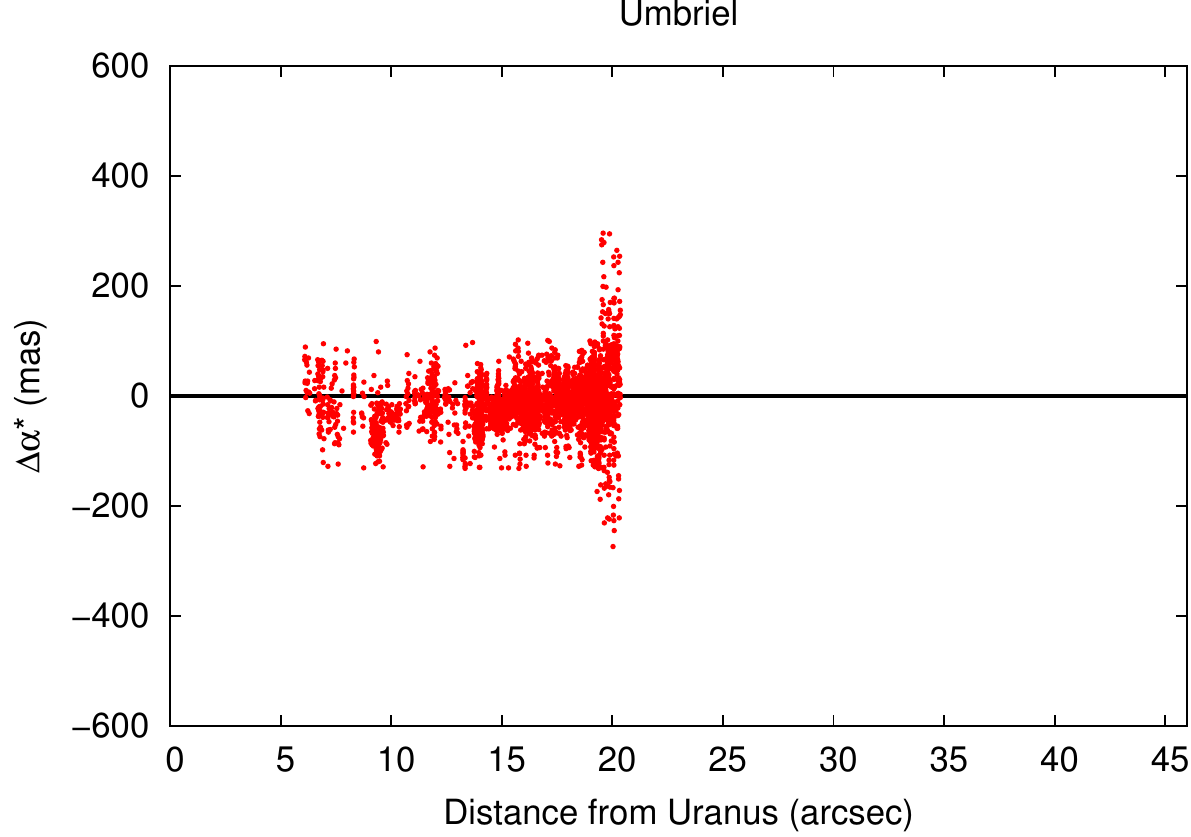}\includegraphics[scale=0.5]{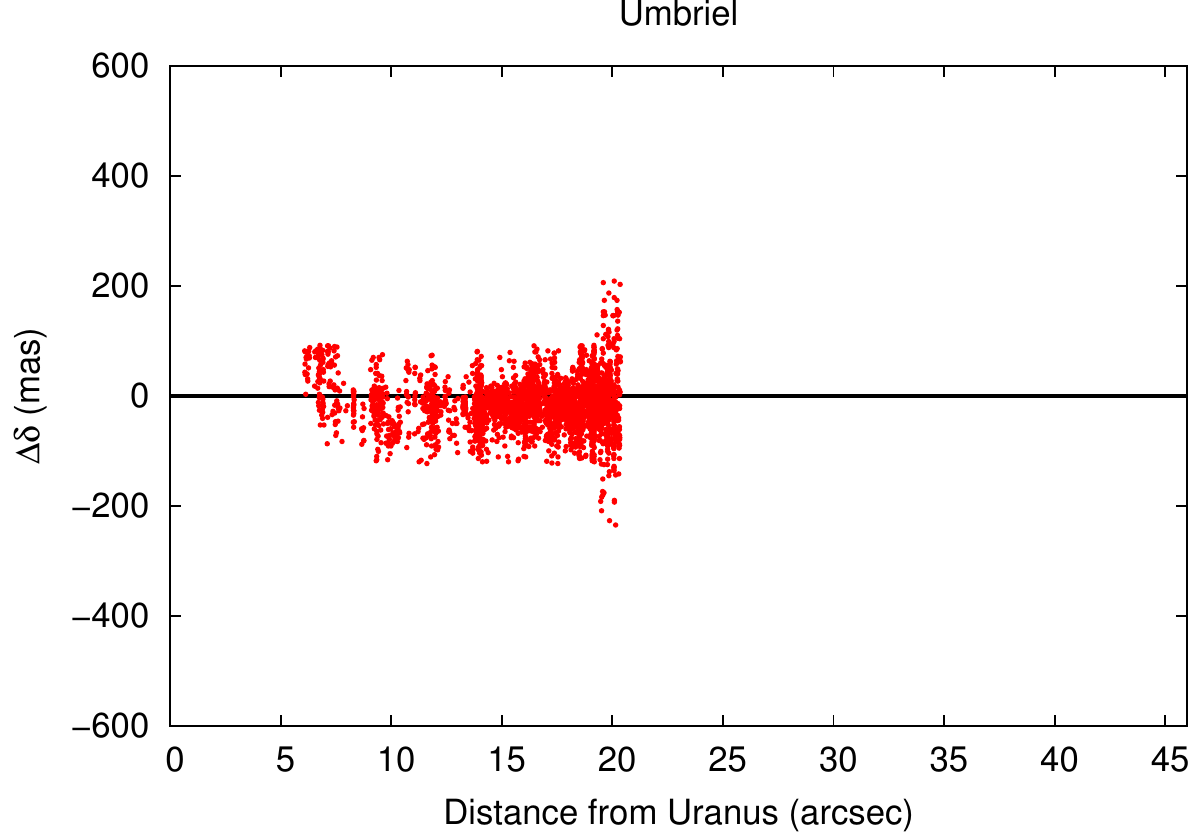}\includegraphics[scale=0.5]{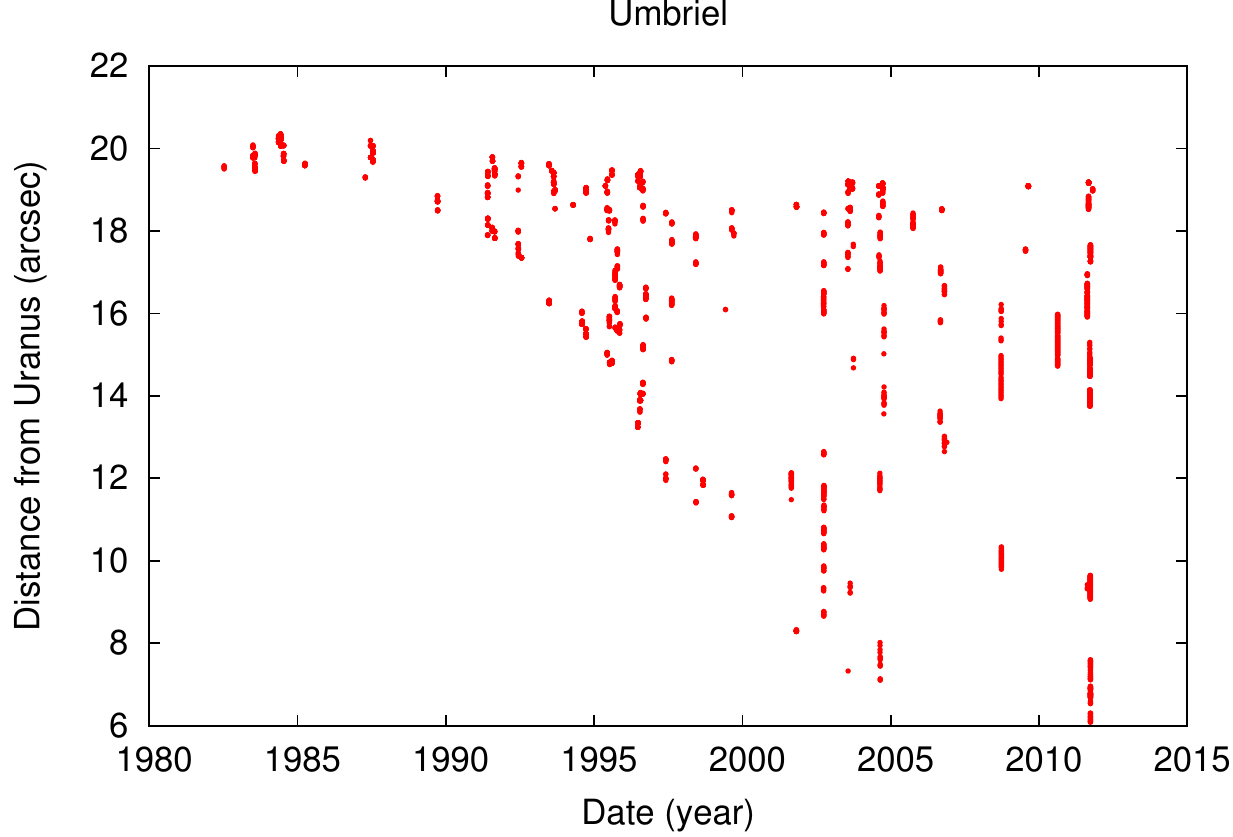}}
\label{fig:umbriel_dist}
\end{figure*}

\begin{figure*}
\caption{Same as Fig.~\ref{fig:miranda_dist} for Titania.}
{\includegraphics[scale=0.5]{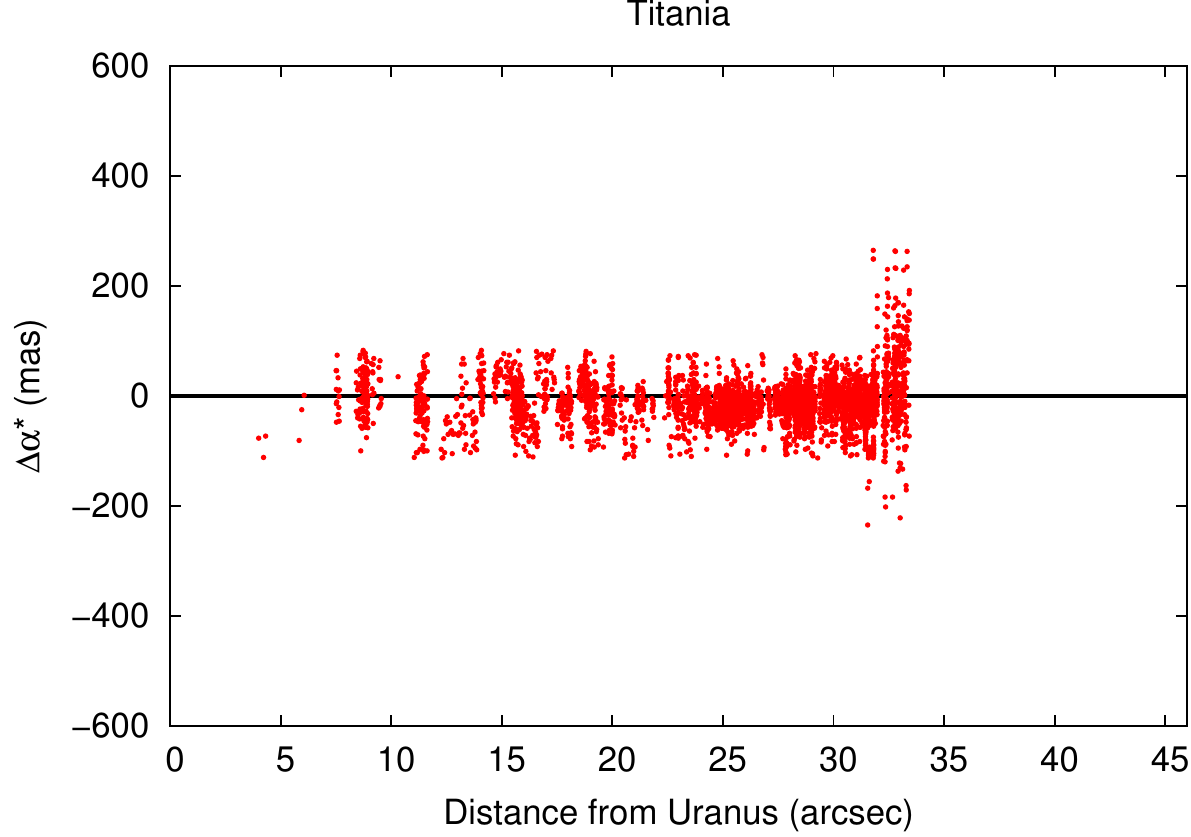}\includegraphics[scale=0.5]{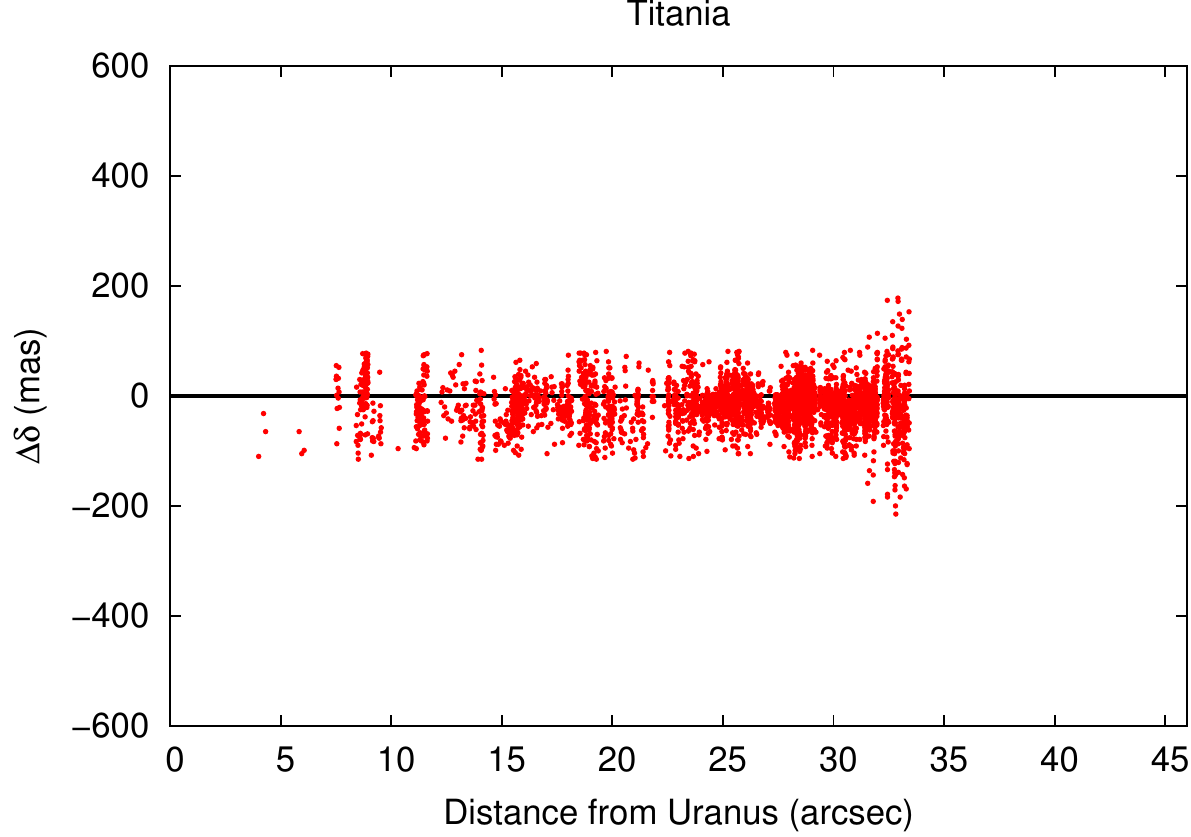}\includegraphics[scale=0.5]{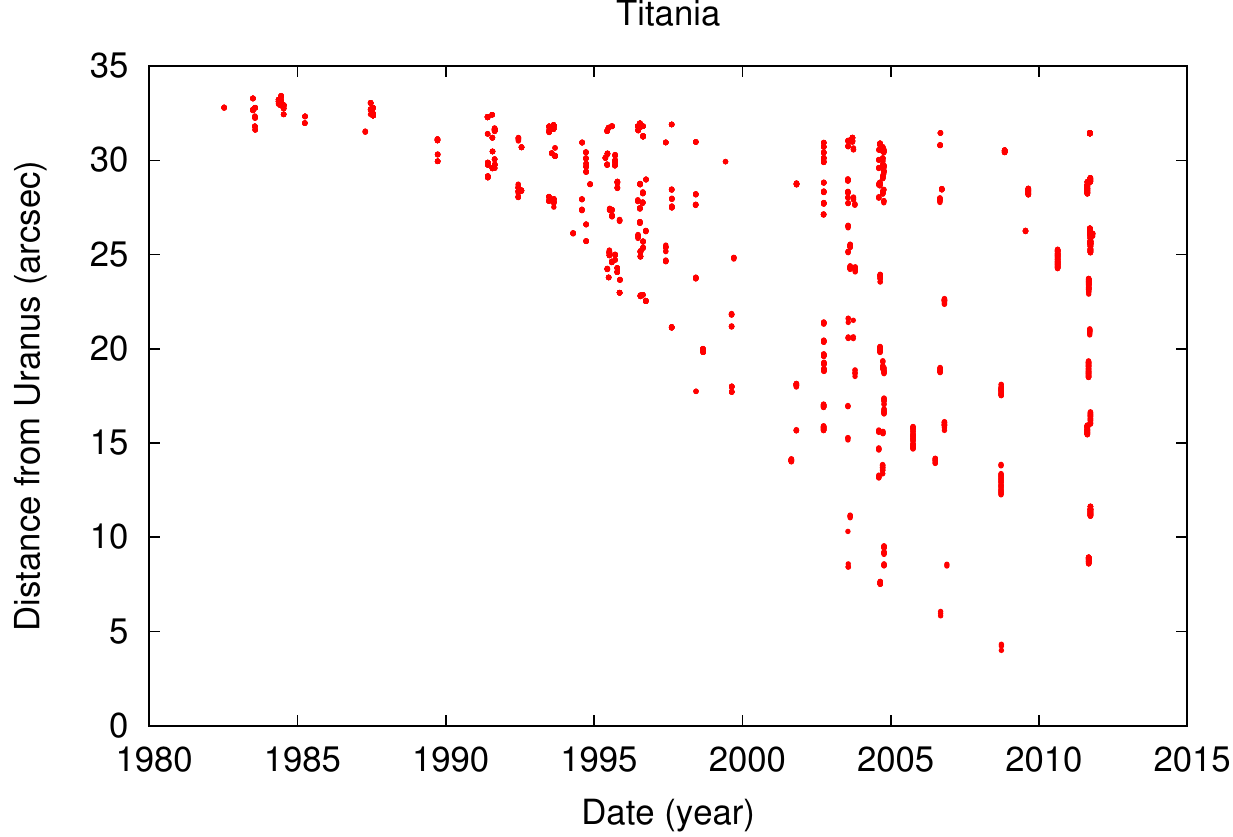}}
\label{fig:titania_dist}
\end{figure*}

\begin{figure*}
\caption{Same as Fig.~\ref{fig:miranda_dist} for Oberon.}
{\includegraphics[scale=0.5]{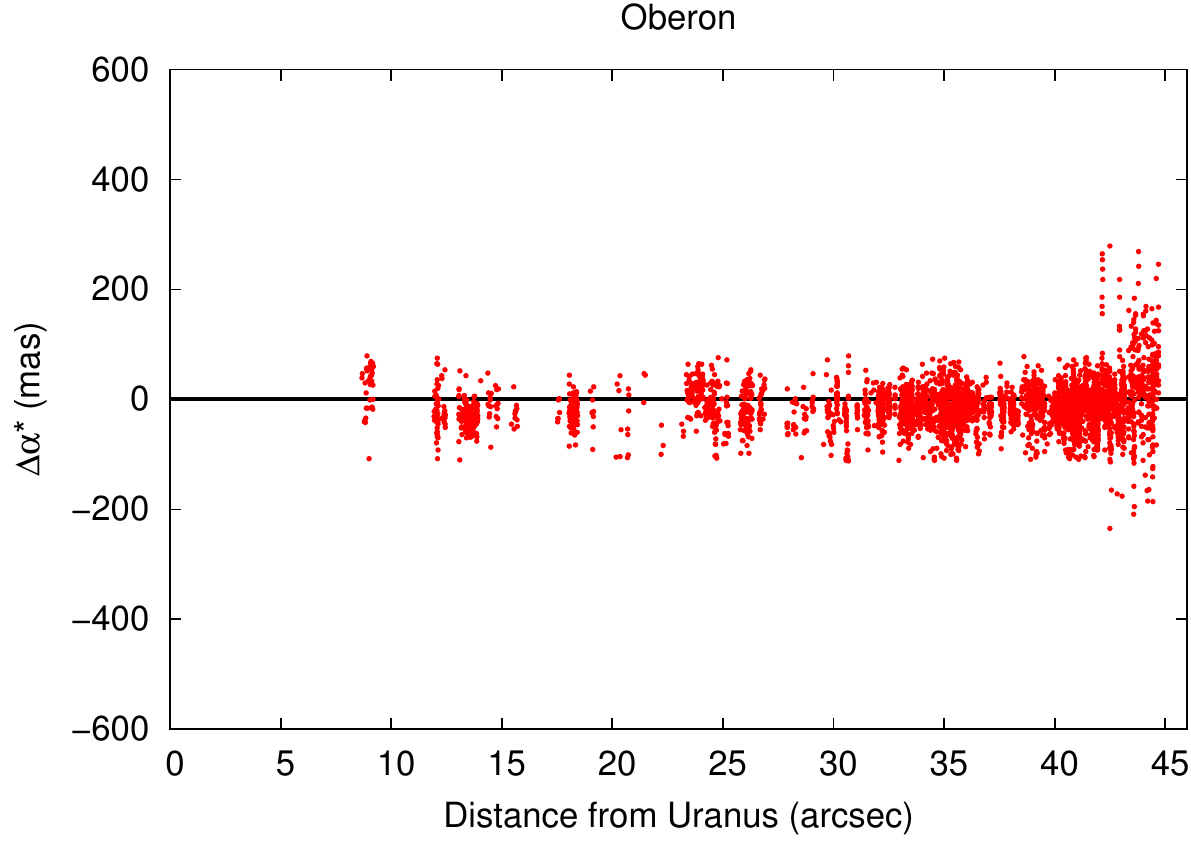}\includegraphics[scale=0.5]{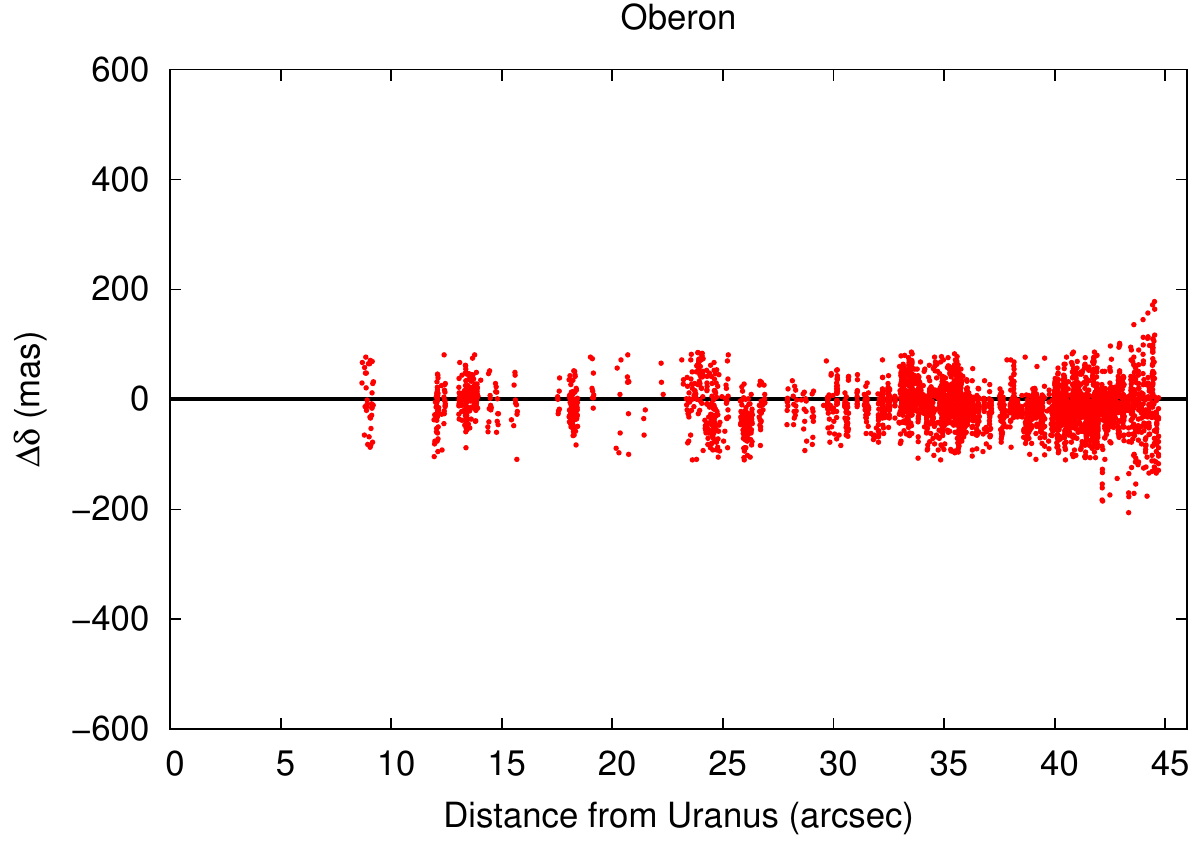}\includegraphics[scale=0.5]{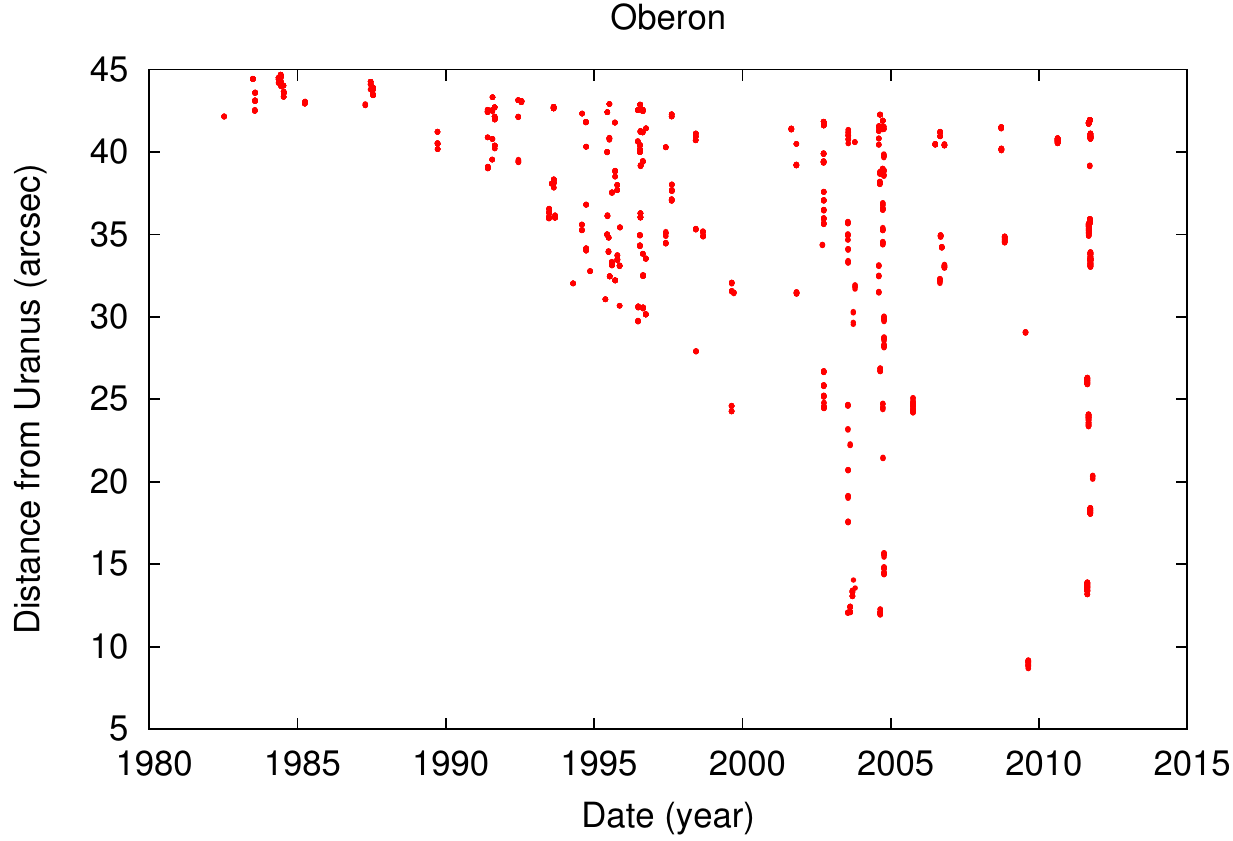}}
\label{fig:Oberon_dist}
\end{figure*}

\begin{table}
\centering
\caption[Observation intervals.]{Time intervals that encompass the observations. CCD (1): data from \citet{2003AJ....125.2714V}. CCD (2): data from C15. Note: photographic plates were used until 1988 but there were no observations of the main satellites of Uranus made in that year.}
\begin{tabular}{cc}
\hline
Data set & Time interval \\
 &  \\
\hline
Photographic plates  & 28MAY1982 to 19JUL1987  \\
 CCD (1)             & 18SEP1989 to 23AUG2004  \\
 CCD (2)             & 09JUN1992 to 25OCT2011  \\
\hline
\end{tabular}
\label{tab:timeobsinterval}
\end{table}

\begin{figure}
\caption{Average number of reference stars per year. These data were determined from the set of observations of Oberon}.
\centerline{\includegraphics[scale=0.5]{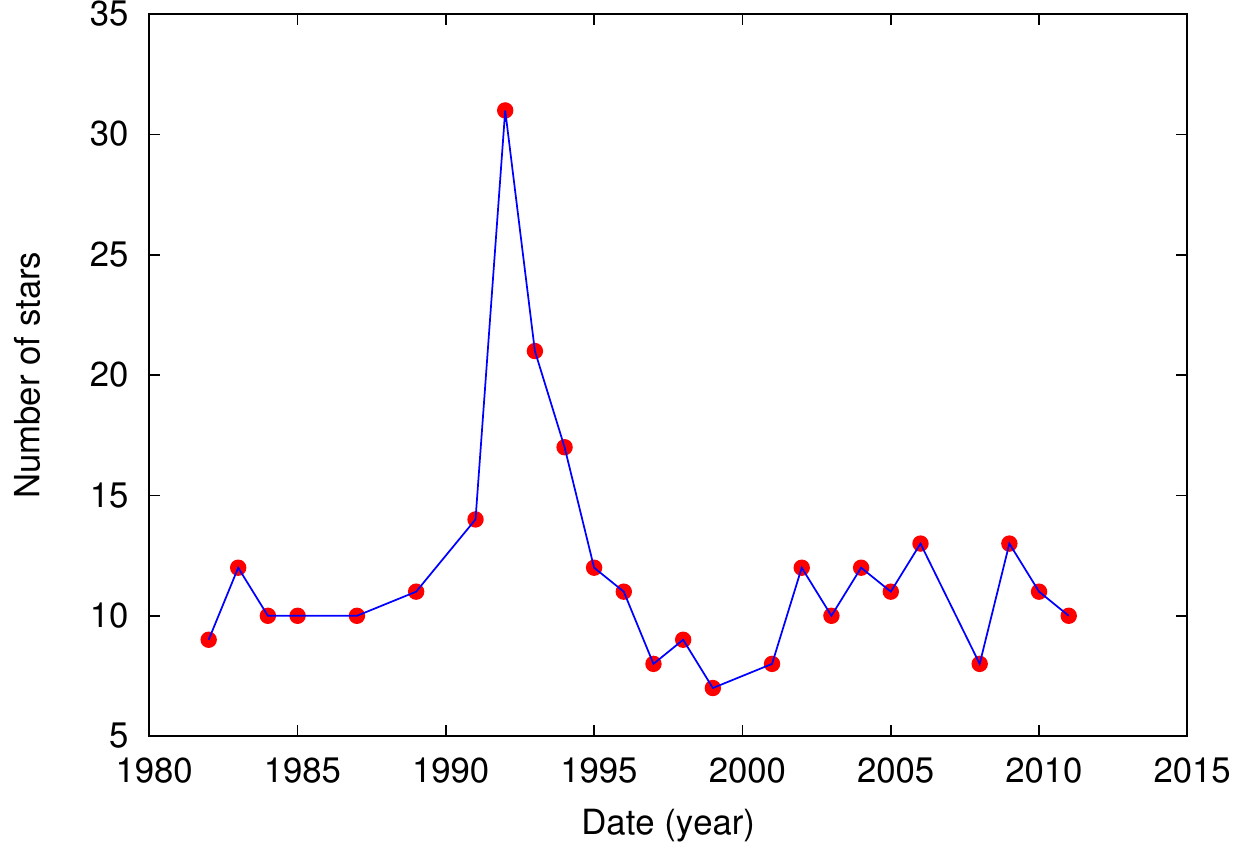}}
\label{fig:refstaryear}
\end{figure}


\begin{table}
\centering
\caption[Overall accuracies and offsets for DE435.]{olumns: satellite name; offsets (observed minus ephemerides - DE435+ura111 and DE435+NOE-7-2013-MAIN) in right ascension and declination for plate and CCD measurements. Standard deviations are given between parenthesis. Last column shows the total number of plate and CCD filtered measurements, respectively, for each satellite. The notation "$<>$" indicates mean value. The "*" means multiplication by the cosine of the declination. Standard deviations are those of the measurements, not of the mean}.
\begin{tabular}{cccccc}
\hline

 & \multicolumn2c{ura111} & \multicolumn2c{NOE-7-2013-MAIN} & Number of \\
Satellite & $<\Delta\alpha*>$ & $<\Delta\delta>$ & $<\Delta\alpha*>$ & $<\Delta\delta>$ & objects\\
 & \multicolumn2c{(mas)} & \multicolumn2c{(mas)} & \\
\hline
Miranda  & $-9 (\pm96)$ & $-20 (\pm73)$ & $-9 (\pm95)$ & $-22 (\pm76)$ & 1680 \\
Ariel  & $-9 (\pm48)$ & $-10 (\pm39)$ & $-8 (\pm48)$ & $-10 (\pm39)$ & 3485 \\
Umbriel  & $-12 (\pm49)$ & $-15 (\pm42)$ & $-12 (\pm49)$ & $-15 (\pm42)$ & 3666 \\
Titania  & $-11 (\pm41)$ & $-16 (\pm37)$ & $-10 (\pm42)$ & $-16 (\pm36)$ & 4442 \\
Oberon  & $-14 (\pm40)$ & $-13 (\pm37)$ & $-13 (\pm40)$ & $-13 (\pm37)$ & 4456 \\
\hline
\end{tabular}
\label{tab:overallaccuracy}
\end{table}

\begin{table*}
\tiny\tiny
\centering
\caption[Overall accuracies and offsets.]{Columns: satellite name; differences in right ascension and declination between the offsets of a satellite (except Oberon) and those of Oberon in the sense Satellite minus Oberon as a function of the separation between the satellite and Uranus. This distance is given by {\it d}. No differences were formed when the distance between Oberon and Uranus was smaller than $25^{\prime\prime}$. Standard deviations are given between parenthesis. The notation "$<>$" indicates mean value. The distance range $0^{\prime\prime}\leq d\leq 5^{\prime\prime}$ presented 0 measurement to Miranda, Ariel and Umbriel and 3 to Titania}. The symbol "$\Delta\Delta$" indicates differences between offsets. The "*" means multiplication by the cosine of the declination. Standard deviations are those of the measurements, not of the mean.
\begin{tabular}{cccccccccccccc}
\hline
Sat. & $<\Delta\Delta\alpha*>$ & $<\Delta\Delta\delta>$ & $<\Delta\Delta\alpha*>$ & $<\Delta\Delta\delta>$ & $<\Delta\Delta\alpha*>$ & $<\Delta\Delta\delta>$ & $<\Delta\Delta\alpha*>$ & $<\Delta\Delta\delta>$ & $<\Delta\Delta\alpha*>$ & $<\Delta\Delta\delta>$ & $<\Delta\Delta\alpha*>$ & $<\Delta\Delta\delta>$ \\
 & \multicolumn2c{(mas)} & \multicolumn2c{(mas)} & \multicolumn2c{(mas)} & \multicolumn2c{(mas)}  & \multicolumn2c{(mas)} & \multicolumn2c{(mas)}  \\
 & \multicolumn2c{($5^{\prime\prime}\leq d\leq 10^{\prime\prime}$)} & \multicolumn2c{($10^{\prime\prime}\leq d\leq 15^{\prime\prime}$)} & \multicolumn2c{($15^{\prime\prime}\leq d\leq 20^{\prime\prime}$)} & \multicolumn2c{($20^{\prime\prime}\leq d\leq 25^{\prime\prime}$)} & \multicolumn2c{($25^{\prime\prime}\leq d\leq 30^{\prime\prime}$)} & \multicolumn2c{($30^{\prime\prime}\leq d\leq 35^{\prime\prime}$)} \\
\hline
Miranda  & $-2 (\pm84)$ & $-5 (\pm69)$  
         & $-$ & $-$
         & $-$ & $-$ 
         & $-$ & $-$
         & $-$ & $-$
         & $-$ & $-$ 
         \\
Ariel    & $+14 (\pm40)$ & $+8 (\pm39)$  
         & $+3 (\pm37)$ & $-1 (\pm31)$
         & $-$ & $-$ 
         & $-$ & $-$
         & $-$ & $-$
         & $-$ & $-$ 
         \\
Umbriel  & $+18 (\pm39)$ & $+9 (\pm45)$  
         & $-4 (\pm24)$ & \hskip 3pt $0 (\pm33)$
         & $+3 (\pm37)$ & $-2 (\pm33)$ 
         & $-8 (\pm73)$ & $+2 (\pm57)$
         & $-$ & $-$
         & $-$ & $-$ 
         \\
Titania  & $+31 (\pm40)$ & $+8 (\pm42)$  
         & $+11 (\pm30)$ & $-13 (\pm31)$
         & $+5 (\pm37)$ & $+7 (\pm30)$ 
         & $-1 (\pm27)$ & $+1 (\pm29)$
         & $+3 (\pm23)$ & $-5 (\pm29)$
         & $+2 (\pm35)$ & $-4 (\pm35)$ 
         \\
\hline
\end{tabular}
\label{tab:offsetsanddistance}
\end{table*}

\begin{table}
\centering
\caption[Overall accuracies and offsets w.r.t Oberon.]{Satellite mean offsets with respect to Oberon. The "*" means multiplication by the cosine of the declination. The notation "$<>$" indicates mean value. The symbol "$\Delta\Delta$" indicates differences between offsets. Standard deviations are those of the measurements, not of the mean. No restriction to the distance Oberon-Uranus was made}.
\begin{tabular}{cccc}
\hline
Satellite & $<\Delta\Delta\alpha*>$ & $<\Delta\Delta\delta>$ & Number of\\
 & \multicolumn2c{(mas)} & objects\\
\hline
Miranda  & $-2 (\pm84)$ & $-5 (\pm69)$ & 1567 \\
Ariel  & $+4 (\pm38)$ & $+1 (\pm33)$ & 3192 \\
Umbriel  & $0 (\pm37)$ & $-1 (\pm35)$ & 3332 \\
Titania  & $+3 (\pm31)$ & $-3 (\pm33)$ & 4028 \\
\hline
\end{tabular}
\label{tab:refoberon}
\end{table}

\section{Comparison with stellar occultation results and different ephemerides}

Here, we compare our observed positions with those given by stellar occultations and different ephemerides: the
planetary ephemerides DE435\footnote{\url{https://naif.jpl.nasa.gov/pub/naif/generic_kernels/spk/planets/}, see also \url{ftp://ssd.jpl.nasa.gov/pub/eph/planets/ioms/}} (JPL) and INPOP19a\footnote{\url{https://www.imcce.fr/recherche/equipes/asd/inpop/download19a}}~\citep[][]{2019NSTIM.109.....F} (Paris Observatory), and ura111\footnote{\url{https://naif.jpl.nasa.gov/pub/naif/generic_kernels/spk/satellites/}} and NOE-7-2013-MAIN\footnote{\url{ftp.imcce.fr/pub/ephem/satel/NOE/URANUS/SPICE/}} (ephemerides for the motions of the main satellites around the Uranian barycenter). NOE-7-2013-MAIN is a more recent adjustment of the motions of the satellites as compared to that given by \citep{2008P&SS...56.1766L} and ura111 is the latest JPL ephemeris for the Uranian satellites.
INPOP21a~\citep[][]{INPOP21A} is also used for comparisons. This new planetary ephemeris is an update of INPOP19a including additional Juno and Mars orbiters observations, but also including the observations presented in this paper in its construction. In the same way, the recent JPL planetary and lunar ephemerides DE440\footnote{\url{https://naif.jpl.nasa.gov/pub/naif/generic_kernels/spk/planets/}}~\citep[][]{2021AJ....161..105P} is used in the comparisons. 

\subsection{Comparisons with results from stellar occultations}

\citet{2009Icar..199..458W} determined an accurate position of Titania from a stellar occultation on 2001-SEP-08. The difference between that position and DE405\footnote{ftp://ssd.jpl.nasa.gov/pub/eph/planets/ioms/de405.iom.pdf and https://naif.jpl.nasa.gov/pub/naif/generic\_kernels/spk/planets/a\_old\_versions/}+ura027\footnote{https://naif.jpl.nasa.gov/pub/naif/generic\_kernels/spk/satellites/a\_old\_versions/} (based on the GUST86 theory), in the sense
observed minus ephemeris, is

\begin{equation}
\label{id:offw1}
    \begin{cases}
      \Delta\alpha{\rm cos}\delta & = -108 \pm 13 {\rm~mas} \\
      \Delta\delta & = -62 \pm 7{\rm~mas}
    \end{cases}
\end{equation}

We do not have observations in September 2001, but we do have data taken in August and October of that year. In this way, differences were calculated with respect to DE405+ura027 and mean offsets in right ascension and declination were determined from 70 measurements. They are

\begin{equation}
\label{id:offc1}
    \begin{cases}
      \Delta\alpha{\rm cos}\delta & = -95 \pm 32 {\rm~mas} \\
      \Delta\delta & = -123 \pm 35 {\rm~mas}
    \end{cases}
\end{equation}

A significant disagreement can be seen in declination. It is interesting to note, however, that \citet{2009Icar..199..458W} mention in their paper a survey made with the Bordeaux meridian transit circle where it is shown that Uranus' offset, averaged over several months around September 2001 amounts to $\Delta\alpha{\rm cos}\delta=-98\pm 10$ mas and $\Delta\delta=-122\pm 10$ mas \citep[see also][]{refId0}. Note that these offsets are consistent to those given by Eq.~\ref{id:offc1}. 
This is an indication that most of the offsets we find may be attributed to the motion of the barycenter of the Uranian system around that of the Solar System.

A second occultation event by Titania, in 2003-AUG-01 is also reported by \citet{2009Icar..199..458W}, whose offsets are
\begin{equation}
\label{id:offw2}
    \begin{cases}
      \Delta\alpha{\rm cos}\delta & = -127 \pm 20 {\rm~mas} \\
      \Delta\delta & = -97 \pm 13 {\rm~mas}
    \end{cases}
\end{equation}

Our 36 observations of Titania, taken in August 2003 and compared to DE405+ura027 give:

\begin{equation}
\label{id:offc2}
    \begin{cases}
      \Delta\alpha{\rm cos}\delta & = -122 \pm 47 {\rm~mas} \\
      \Delta\delta & = -129 \pm 43 {\rm~mas}
    \end{cases}
\end{equation}

Both results are consistent within their respective $1\sigma$ uncertainties. These comparisons - at least to the positions of Titania used in this section - indicate the consistency of our positions with respect to independent accurate astrometry of the Uranian system as well as the reliability of our standard deviations.

\subsection{Comparisons with different ephemerides}

Here, we present comparisons with the dynamical model for the motion of the five main satellites of Uranus NOE-7-2013-MAIN and ura111, and the planetary ephemerides INPOP19a, INPOP21a, DE435 and DE440.

\subsubsection{DE435$+$ura111 and DE435+NOE-7-2013-MAIN}

Figures~\ref{fig:miranda} to \ref{fig:oberon} show, for right ascension and declination, the behaviour of the offsets (those that survived to the filtering) in the sense observation minus ephemerides as a function of time, true anomaly, and the angle Sun-Observer-Satellite. Those figures also
show that the values presented in Table~\ref{tab:overallaccuracy} are a suitable global representation of the offsets along the x-axes (time, true anomaly, satellite elongation).
 
It should be noted that, as compared to C15, this work presents smaller offsets (absolute values), indicating a better agreement between observations and DE435+ura111. Given that DE432 (used in C15) and DE435 do not differ significantly (few mas level) for Uranus within the time span of our observations, this better agreement comes from the use of Gaia EDR3 (see Table~\ref{tab:overallaccuracy}).

As expected from C15, also given by Table~\ref{tab:overallaccuracy}, NOE-7-2013-MAIN and ura111 yielded very similar results.  This reinforces the indication that most of the offsets we find in the position of the satellites come from the motion of the Uranian barycenter around the barycenter of the Solar System as given by the planetary ephemerides.

Another important feature also comes from the use of Gaia EDR3. Although two images of the Uranian system, taken at sufficiently distant instants, will not have the same reference stars, we expect that those two sets of stars locally materialize accurately the same coordinate axes, namely, the ICRF. In this context, we understand that our observed positions have a better coherence with respect to those from C15 and that this is reflected in the overall offsets and standard deviations as given by Table~\ref{tab:overallaccuracy}.

Table~\ref{tab:offsetsanddistance} complements the information given by Tables~\ref{tab:overallaccuracy} and ~\ref{tab:refoberon} by providing some numerical details about remaining effects of the distance Uranus-satellite on our positions. Figures~\ref{fig:miranda_dist} to \ref{fig:Oberon_dist} illustrate the details of Table~\ref{tab:offsetsanddistance}, whereas Fig.~\ref{fig:refstaryear} estimates the average number of reference stars per year. Overall, they show the effectiveness of the digital coronography procedure to attenuate the effects of the scattered light from Uranus as well as the quality of Gaia EDR3 to materialize the celestial frame even when few reference stars are available.

It should be noted in Table~\ref{tab:offsetsanddistance} that the difference between offsets $\Delta\Delta\alpha*$ for Titania within the distance interval from Uranus $5^{\prime\prime}\leq d\leq 10^{\prime\prime}$, $+31(\pm40)$ mas, is larger than that for the other satellites. Most of the data in this bin of distances come from 81 observations of Titania made in 2011 (the remaining 16 measurements within this same bin of distances from Uranus show values $\Delta\Delta\alpha*$ and $\Delta\Delta\delta$ of few mas only), where an asymmetric light distribution made it difficult to free Titania from the scattered light from Uranus. Given our successful efforts to attenuate that scattered light, and knowing that our positions will
mostly serve to improve the orbits of Uranus and its main satellites, we believe it is useful to keep in our final data set all measurements and to also provide in our catalogue of positions, as mentioned earlier, the distances between the satellites and Uranus.

Table~\ref{tab:refoberon} compares the offsets in right ascension and declination of a given satellite to those of Oberon. The standard deviations derived from these differences between offsets provide a more trustworthy evaluation of angle measurements among point-like objects. These results are similar to those presented in C15 as zero-point errors are cancelled out when differences between offsets are formed. If we consider only measurements from CCD data, then the standard deviations shown in Table~\ref{tab:refoberon} become slightly smaller to Miranda.

\subsubsection{INPOP19a, DE440 and INPOP21a}

In these comparisons, the ephemeris of the five main Uranian moons is still ura111 but the motion of the Uranian barycenter around the barycenter of the Solar System is given by the IMCCE planetary ephemerides INPOP19a and INPOP21a, and the JPL planetary ephemeris DE440 - see Table~\ref{tab:compin19a21ade440} and compare it to Table ~\ref{tab:overallaccuracy}. 
In order to estimate the impact of the Gaia EDR3 reduction on the Uranian barycentric orbits, an update of the INPOP planetary ephemerides has been made by introducing into the INPOP adjustment the Uranian satellite positions presented in this work. In Table~\ref{tab:compin19a21ade440} one can see the clear reduction of the offset in declination of about 30 mas thanks to the use of Gaia EDR3 reduced satellite positions. This result enhances the importance of such determination for the link between planetary planes and the Gaia reference frame.


\begin{table*}
\centering
\caption[Overall accuracies and offsets for INPOP19a and INPOP21a.]{ Columns: satellite name; offsets (observed minus ephemeris) in right ascension and declination for plate and CCD measurements. Standard deviations are given between parenthesis. Last column shows the total number of plate and CCD filtered measurements, respectively, for each satellite. The notation "$<>$" indicates mean value. The "*" means multiplication by the cosine of the declination. Standard deviations are those of the measurements, not of the mean}.
\begin{tabular}{cccccccc}
\hline
 & \multicolumn2c{INPOP19a} & \multicolumn2c{INPOP21a} &  \multicolumn2c{DE440} & Number of\\
Satellite & $<\Delta\alpha*>$ & $<\Delta\delta>$ & $<\Delta\alpha*>$ & $<\Delta\delta> $ & $<\Delta\alpha*>$ & $<\Delta\delta>$ & objects\\
 & \multicolumn2c{(mas)} & \multicolumn2c{(mas)} & \multicolumn2c{(mas)} &\\
\hline
Miranda  & $-14 (\pm94)$ & $-46 (\pm73)$ & $-11 (\pm101)$ & $-16 (\pm74)$& $-15 (\pm96)$ & $-19 (\pm74)$  & 1680 \\
Ariel  & $-3 (\pm48)$ & $-32 (\pm42)$ & $-3 (\pm50)$&$-4 (\pm39)$ &  $-19 (\pm49)$ & $-9 (\pm39)$ & 3485 \\
Umbriel  & $-6 (\pm48)$ & $-37 (\pm44)$ & $-7 (\pm50)$ & $-9 (\pm42)$ & $-22 (\pm51)$ & $-14 (\pm42)$ & 3666 \\
Titania  & $-2 (\pm41)$ & $-38 (\pm39)$ & $-3 (\pm42)$ & $-9 (\pm37)$  & $-21 (\pm43)$ & $-16 (\pm37)$ & 4442 \\
Oberon  & $-5 (\pm40)$ & $-34 (\pm39)$ & $-6 (\pm41)$ & $-5 (\pm38)$ & $-24 (\pm42)$ & $-12 (\pm37)$ & 4456 \\
\hline
\end{tabular}
\label{tab:compin19a21ade440}
\end{table*}

The planetary and lunar ephemerides DE440 has recently replaced\footnote{\url{https://ssd.jpl.nasa.gov/?horizons_news}} DE430 and includes seven years of new data in its computation, in addition to using improved models and data calibration \citep[][]{2021AJ....161..105P}. Although the main objective of this paper is to present the astrometry of the main Uranus' satellites and its quality, accomplished with the ephemerides used so far, it is interesting to have a comparison between our positions and those from DE440. This is also given by Table~\ref{tab:compin19a21ade440}.

The results presented in Table~\ref{tab:overallaccuracy} for ura111 differ from those in Table~\ref{tab:compin19a21ade440} only by the use of the planetary ephemerides (DE435 in the first table and INPOP19a, INPOP21a and DE440 in the latter). Both tables lead to the same conclusion: with respect to our observations, a systematic effect in the motion of the Uranian barycenter around the Solar System barycenter can be seen.

\subsection{The catalogue}

The positions of the Uranian main satellites determined here are provided in the form of catalogues and are available at CDS via anonymous ftp to cdsarc.u-strasbg.fr (130.79.128.5). An extract with the first lines of the catalogue for Oberon is shown by Table~\ref{tab:catoberon}. 

All right ascensions and declinations in the catalogue are astrometric ones, with origin at the observing site (IAU code: 874) and referred to the ICRF. We also emphasize that this work does not provide positions of Uranus and does not make any direct measurement of them. All positions in the catalogue are obtained from images containing the Uranian satellites and stars for which astrometry can be extracted from Gaia EDR3 (reference stars). Calculated positions of the satellites were obtained from a planetary ephemeris plus
an ephemeris that describes the motion of the satellites around the Uranian barycenter.

\begin{table}
\centering
\caption[Catalogue extract for Oberon.]Angular measurements have the same meaning as those provided by the the JPL's HORIZONS system (ephemeris service). Columns: astrometric right ascension and declination, with origin at the observing site (IAU code 874) and referred to the ICRF; Year,  Month, Day and Fraction of a day (YMDF) along with JD of the observation mid-time; satellite apparent position, in the plane of the sky, with respect to the central body in the sense satellite minus central body. The difference in right ascension is multiplied by the cosine of the declination.
\hskip 10pt
\begin{tabular}{cccccc}
\hline
RA & DEC & YMDF & JD & DX & DY\\
\multicolumn2c{(ICRF)} & \multicolumn2c{(UTC)} & \multicolumn2c{(arcsec)}\\
\hline
15 56 10.5902 & $-$20 15 08.385 & 19820711.94166667 & 2445162.44166667 & $-35.875$ & $-22.175$ \\
15 56 10.5410 & $-$20 15 08.224 & 19820711.94930556 & 2445162.44930556 & $-35.953$ & $-22.039$ \\
15 56 10.4777 & $-$20 15 07.841 & 19820711.95902778 & 2445162.45902778 & $-36.052$ & $-21.865$ \\
15 56 10.4221 & $-$20 15 07.626 & 19820711.96562500 & 2445162.46562500 & $-36.118$ & $-21.747$ \\
15 56 10.3844 & $-$20 15 07.397 & 19820711.97326389 & 2445162.47326389 & $-36.195$ & $-21.610$ \\
15 56 10.3377 & $-$20 15 07.192 & 19820711.97916667 & 2445162.47916667 & $-36.254$ & $-21.504$ \\
15 56 10.2973 & $-$20 15 06.912 & 19820711.98541667 & 2445162.48541667 & $-36.316$ & $-21.392$ \\
\vdots & & & & \\
\hline
\end{tabular}
\label{tab:catoberon}
\end{table}

\begin{table}
\centering
\caption[Weighing suggestions.]{Columns: satellite name; standard deviations, in right ascension and declination, that can be used as weights for each measurement in position; total number of observations involved in each determination of standard deviations. The "*" means multiplication by the cosine of the declination}.
\begin{tabular}{cccc}
\hline
Satellite & $\sigma_{\alpha}$* & $\sigma_{\delta}$ & Number of\\
 & \multicolumn2c{(mas)} & objects\\
\hline
Miranda  & 96 & 73 & 1680\\
Ariel/Umbriel/Titania/Oberon   & 44 & 39 & 16049\\
\hline
\end{tabular}
\label{tab:accuracies4all}
\end{table}

\section{Conclusions and Comments}

We provide a set of accurate positions of Uranus' main satellites spread over almost 3 decades of observations with photographic plates and CCDs.

On average, we obtained offsets $\Delta\alpha{\rm cos}\delta=-11~(\pm 52)$ milli-arcseconds and $\Delta\delta=-14~(\pm 43)$ milli-arcseconds for the differences in the sense observation minus ephemerides (DE435+ura111). When the offsets $\Delta\alpha{\rm cos}\delta$ and $\Delta\delta$ of Miranda, Ariel, Umbriel and Titania are compared, for the same dates, to those of Oberon, we obtain overall differences of $+2~(\pm 45)$ milli-arcseconds in right ascension and $-2~(\pm 40)$ milli-arcseconds in declination. Comparisons of the observations with various ephemerides, along with results from stellar occultations, indicate a possible offset in the (Solar System) barycentric position of the Uranian system barycenter.
These results are an improvement with respect to our previous works, individually, not only because of the longer time span of the observations but mainly due to the use of Gaia EDR3 as reference for astrometry, which decreased the systematic effects in the observed positions of the satellites.

The data presented here will mostly serve as a source to improve the orbits of the satellites and that of Uranus. In this context, a number that could describe the accuracies of the satellites' positions, in a general way, is useful. The standard deviations in right ascension and declination, from the full set of observations, is a simple way to provide conservative figures and are shown in Table~\ref{tab:accuracies4all}. We recognize, however, that the observations of Miranda are not as accurate as those of the other satellites so we provide them separately.

\section*{Acknowledgements}
The following authors acknowledge the respective CNPq grants: J.I.B.C. 308150/2016-3 and 305917/2019-6; RV-M 304544/2017-5, 401903/2016-8; M.A 427700/2018-3, 310683/2017-3 and 473002/2013-2. The results were based on observations taken at Pico dos Dias Observatory of the National Laboratory of Astrophysics (LNA/Brazil).
Software Routines from the IAU SOFA Collection were used. Copyright \textcopyright~International Astronomical Union Standards of Fundamental Astronomy (\url{http://www.iausofa.org}). The authors acknowledge the numerous and dedicated colleagues that contributed to the observing runs over many years.

\bibliographystyle{aa}
\bibliography{aa.bib} 
\end{document}